\documentclass[twocolumn]{aastex63}
\usepackage{amsmath}
\usepackage{lineno}

\received{20 October 2021}
\revised{26 January 2022}
\accepted{16 February 2022}

\accepted{\textit{Astrophysical Journal}}
\shorttitle{Internal forcing on irradiated giants}
\shortauthors{Lian et al.}

\begin{document}
\title{Influences of internal forcing on atmospheric circulations of irradiated giant planets\footnote{Released on xxx}}

\correspondingauthor{Yongyun Hu}
\email{yyhu@pku.edu.cn}

\author{Yuchen Lian}
\affiliation{Laboratory for Climate and Ocean-Atmosphere Studies, Department  of  Atmospheric  and  Oceanic  Sciences,  Peking
University, Beijing 100871, China}
\author{Adam P. Showman}
\affiliation{Laboratory for Climate and Ocean-Atmosphere Studies, Department  of  Atmospheric  and  Oceanic  Sciences,  Peking
University, Beijing 100871, China}
\affiliation{Lunar and Planetary Laboratory, University of Arizona, 1629 University Boulevard, Tucson AZ 85721, USA}
\author{Xianyu Tan}
\affiliation{Atmospheric Oceanic and Planetary Physics, Department of Physics, University of Oxford OX1 3PU, UK}
\author{Yongyun Hu}
\affiliation{Laboratory for Climate and Ocean-Atmosphere Studies, Department  of  Atmospheric  and  Oceanic  Sciences,  Peking
University, Beijing 100871, China}

\begin{abstract}
Close-in giant planets with strong stellar irradiation show atmospheric circulation
patterns with strong equatorial jets and global-scale stationary waves. So far, almost all modeling works on atmospheric circulations of such giant planets have mainly considered external radiation alone, without taking into account the role of internal heat fluxes or just treating it in very simplified ways. Here, we study atmospheric circulations of strongly irradiated giant planets by considering the effect of internal forcing, which is characterized by small-scale stochastic interior thermal perturbations, using a three-dimensional atmospheric general circulation model. We show that the  perturbation-excited waves can largely modify atmospheric circulation patterns in the presence of relatively strong internal forcing. Specifically, our simulations demonstrate three circulation regimes: superrotation regime, midlatitude-jet regime, and quasi-periodic oscillation regime, depending on the relative importance of external and internal forcings. It is also found that strong internal forcing can cause noticeable modifications of the thermal phase curves. 
\end{abstract}

\keywords{exoplanet atmosphere ------ atmospheric circulation ------ brown dwarfs ------ hot Jupiters}

\section{Introduction}
Atmospheric circulations of hot Jupiters have drawn extensive theoretical and numerical studies (see reviews by \citealt{heng-showman-2015} and \citealt{showman2020review}). Both idealized and global three-dimensional (3D) models demonstrated that the strong day-night temperature contrast of hot Jupiters causes a strong equatorial eastward jet stream, i.e., the equatorial superrotation \citep{showman-etal-2002, showman-etal-2011, cooper-etal-2005, showman-fortney-etal-2008, thrastarson-etal-2010, liu-etal-2012, showman-lewis-etal-2015,mayne-etal-2017,mendonca-2020}. \citet{showman-etal-2011}, \citet{hu-etal-2013} and \citet{hammond-etal-2018} further showed that the equatorial superrotation is caused by interactions between the mean flow and the stationary waves which are excited by the permanent day-night temperature contrasts. \citet{tsai-etal-2014} and \citet{debras-etal-2020} showed that this mechanism holds in continuously stratified atmospheres. In consequence, the eastward equatorial jet causes a 
eastward shift of the hot spot, which has been confirmed by phase-curve observations \citep{knutson-etal-2007,knutson-etal-2008}. 

However, many previous studies treated the internal heat flux as a uniform heat flux but do not account for the time and space dependent fluctuations that may be caused by the internal heat flux. For most hot jupiters, it is reasonable to neglect internal heat fluxes because the typical stellar radiation received by hot Jupiters is about $10^3$ times stronger than their internal heat fluxes \citep{baraffe-etal-2009}. However, it is speculated that young hot Jupiters may have strong internal heat fluxes \citep{fortney-etal-2010}. Observations showed evidence that the radii of hot Jupiters are larger than we expected, suggesting that the inflated hot Jupiters may have more internal heat fluxes than previously thought \citep[e.g.][]{guillot-etal-2002,thorngren-etal-2019}. Internal heat fluxes strengthen the turbulence, which extends to the radiative–convective boundary and transports external heat into the interior, inflating the planet \citep{youdin-mitchell-2010}. Thus, the internal effective temperature, which is often assumed to be like that of Jupiter ($\sim$ $100$ ${\rm K}$) for many hot Jupiter models, should be much higher for inflated hot Jupiters \citep[e.g.][]{thorngren-etal-2019}, and internal heat fluxes may have important effects on atmospheric circulations of hot Jupiters.

An extreme case of the importance of internal heat fluxes is isolated brown dwarfs, which have about $10^3$ to $10^6$ ${\rm W}$ ${\rm m^{-2}}$ internal heat fluxes due to their large internal energy accreted during their formation, gravitative-contraction and lithium fusion process \citep{burrows-etal-2001,marley-etal-2015}, while external forcing is negligible. The interior heat is transported from the interior to the radiative cooling layer via convection, and convection interacts with the overlying stratified layers to generate waves, as shown in regional 2D models \citep{freytag-etal-2010}. These waves can interact with large-scale flow and drive global-scale circulations, which has been demonstrated by global shallow water models \citep{zhang-etal-2014}. \citet{showman-etal-2013} demonstrated the interaction between global convection and rapid rotation of brown dwarfs leads to symmetrical equator-pole temperature and wind structures near the top of the convective zone. \citet{showman-etal-2019} showed that zonal jets and long-term oscillations in the equatorial stratosphere are ubiquitous for 3D general circulation models (GCM) of brown dwarf-type simulations.

In addition, more than a dozen brown dwarfs have been discovered in close-in orbits around sun-like stars and white dwarfs \citep{siverd-etal-2012, casewell-etal-2012, bayliss-etal-2016}. These irradiated brown dwarfs are also likely tidally locked, with large day-night temperature contrasts. \citet{tan-etal-2020} investigated atmospheric circulations of fast rotating brown dwarfs around white dwarfs, using GCMs with a simple Newtonian cooling scheme. They showed that the fast-rotating irradiated brown dwarfs have stronger day-night temperature contrasts and smaller phase-curve offsetting compared with slow-rotating hot Jupiters, even if they have developed strong equatorial jets. \citet{lee-etal-2020} also simulated the white-dwarf-brown-dwarf system, with a double-grey radiative transfer scheme, similarly showing a large day-night temperature contrast. However, these studies did not consider the role of internal heat fluxes in atmospheric circulations either. Given possible strong internal heat fluxes, the effect of external-internal interactions on atmospheric circulations for these brown dwarfs are of great interests. 

The major purpose of this paper is to study atmospheric circulations of hot jupiters and brown dwarfs that experience both permanent day-to-night irradiation differences and internal heat fluxes, and how atmospheric circulations change under different strengths of the external and internal forcings. In this study, the external forcing is represented by an idealized permanent day-to-night temperature difference, and the internal forcing is parameterized by small-scale stochastic thermal perturbations in deep layers, mimicking effects of convective perturbations on the bottom of the stratified layers \citep{showman-etal-2019}. In addition, changes of atmospheric circulations may alter thermal phase curves of hot Jupiters and brown dwarfs. Observations have shown time variations of thermal phase curves of hot Jupiters, such as Kepler-76b \citep{espinoza-etal-2015, jackson-etal-2019} and brown dwarfs \citep{wilson-etal-2014, artigau-etal-2009, cushing-etal-2016}. Thus, the other purpose of this paper is to investigate the shapes of thermal phase curves influenced by different strengths of external and internal forcings.

The paper is organized as follows. The GCM used here is described in Section \ref{Model}. Simulation results are shown in Section \ref{Result}. Shapes of synthetic thermal phase curves are discussed in Section \ref{Lightcurve}. Discussion and conclusions are summarized in Section \ref{conclusion}.

\section{Model description and Experimental design}
\label{Model}

The 3D GCM used here solves the global hydrostatic primitive equations in pressure coordinates which are described as follows:

\begin{equation}
\label{eqn1}
{\frac{d\textbf{u}}{dt}}=-\nabla_p \Phi - f\textbf{k} \times \textbf{u}-{\frac{\textbf{u}}{\rm \tau_{drag}}},
\end{equation}

\begin{equation}
\label{eqn2}
{\frac{\partial \Phi}{\partial p}}=-{\frac{1}{\rho}},
\end{equation}

\begin{equation}
\label{eqn3}
\nabla_p \cdot \textbf{u}+{\frac{\partial\omega}{\partial p}}=0,
\end{equation}

\begin{equation}
\label{eqn4}
{\frac{dT}{dt}}={\frac{q_1+q_2}{c_{p}}}+{\frac{\omega}{\rho c_{p}}}.
\end{equation}
Equation (\ref{eqn1}) is the horizontal momentum equation, Equation (\ref{eqn2}) is the hydrostatic balance equation, Equation (\ref{eqn3}) is the continuity equation, and Equation (\ref{eqn4}) is the thermodynamics equation. In those equations, $\frac{d}{dt}$ is the material derivative, $\mathbf{u}$ represents horizontal velocity vector, ${\rm \omega}$ is the vertical velocity at pressure coordinates, ${\rm \nabla_p}$ is the horizontal pressure gradient, $f$ ${\rm =2\Omega sin\phi}$ is the Coriolis parameter, ${\rm \Omega}$ is the planetary rotation rate, ${\rm \phi}$ is latitude, $\Phi$ is gravitational potential, $c_p$ is the specific heat at constant pressure and $\mathbf{k}$ is the vertical vector. 

We include a deep frictional drag in our model, similar to that used in \cite{komacek-etal-2016}. The frictional drag linearly relaxes the horizontal velocity towards zero over a characteristic drag timescale $\tau_{\rm drag}$ that decreases with increasing pressure. Drag exists only near the bottom ($> p_{\rm drag}=$ 10 bars and $< p_{\rm bot}=$ 400 bars), representing the interaction between the upper layers and the quiescent interior. $\tau_{\rm drag,bot}$ is 10 Earth days at bottom. $\tau_{\rm drag}$ increases with the decreasing pressure, and reaches infinite when pressure equals $p_{\rm drag}$. (Equation \ref{pdrag})

\begin{equation}
\frac{1}{\tau_{\rm drag}}=\left\{
             \begin{array}{lcl}
0, \quad p < p_{\rm drag} \\
\frac{1}{\tau_{\rm drag,bot}}-\frac{1}{\tau_{\rm drag,bot}}{\frac{log({\frac{p}{p_{\rm bot}}})}{log({\frac{p_{\rm drag}}{p_{\rm bot}}})}}, p\ge p_{\rm drag} \\
              \end{array}
              \right.
\label{pdrag}
\end{equation}

\begin{figure}[!ht]
    \centering
    \includegraphics[height=18.0cm,width=8.5cm]{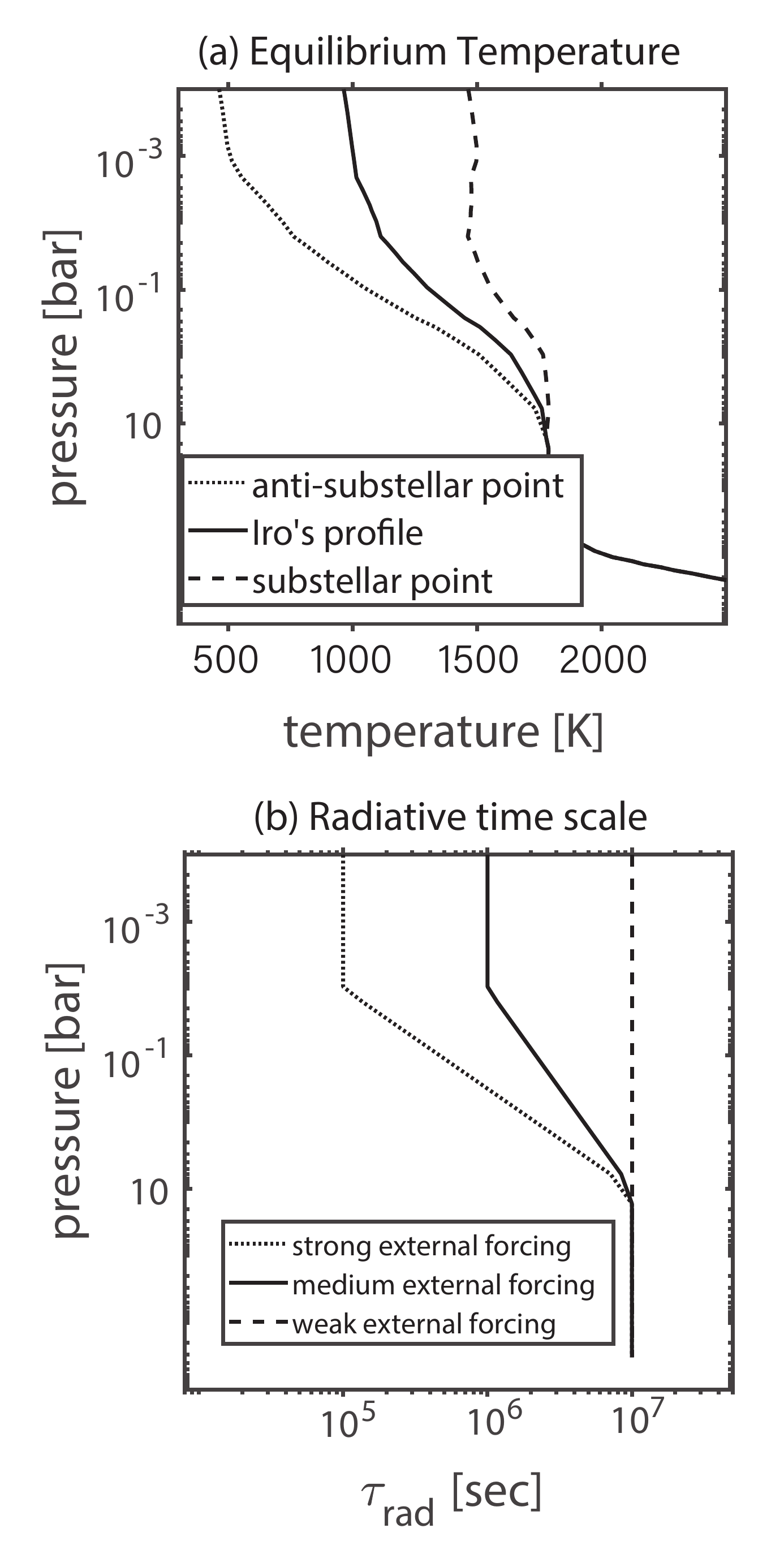}
    \caption{Vertical profiles of equilibrium temperatures and radiative time scales. (a) Equilibrium temperatures, and (b) radiative time scales for different external forcings.}
    \label{fig:1}
\end{figure}

The system is thermally forced by two ways. First, we adopt an idealized Newtonian cooling scheme to represent the permanent day-night temperature contrast that relaxes the 3D temperature field to a prescribed equilibrium day-night profile over characteristic radiative timescales. This scheme has been widely used in previous studies for hot Jupiters \citep{cooper-etal-2005,showman-etal-2008,liu-etal-2012,komacek-etal-2016}.
The heating rate $q_1$ (J ${\rm kg^{-1}}$ ${\rm s^{-1}}$) in Equation (\ref{eqn4}) represents that of the external forcing:
\begin{equation}
{\frac{q_{ 1}}{c_{ p}}}=-{\frac{T(\lambda,\phi,p,t)-T_{\rm eq}(\lambda,\phi,p)}{\tau_{\rm rad}(p)}},
\end{equation}
where ${\rm \lambda}$, ${\rm \phi}$, $p$, $t$ are longitude, latitude, pressure, and time, respectively. Temperature is relaxed to the equilibrium state $T_{\rm eq}$ over a characteristic radiative timescale ${\rm \tau_{rad}}$ that is a function of pressure. The mean equilibrium temperature $T_{\rm eq}$ profile is adopted from \citet{iro-etal-2005} and the global equilibrium temperature field is prescribed as below (also shown as the left panel of Figure \ref{fig:1}):

The reference temperature profile $T_{\rm ref}$ is from \citet{iro-etal-2005} and the mean equilibrium temperature $T_{\rm eq}$ is adopted by the equation prescribed as below (also shown as the left panel of Figure \ref{fig:1}):

\begin{equation}
T_{\rm eq}=\left\{
          \begin{array}{lcl}
    T_{\rm ref}, \quad \mu \leq 0 \\
    T_{\rm ref}+\Delta T_{\rm eq}\mu, & \quad \mu > 0
          \end{array}
        \right.
\end{equation}

where $\mu$ is the cosine of the irradiation angle, and
{
\begin{equation}
\Delta T_{\rm eq}=\left\{
             \begin{array}{lcl}
\Delta T_{\rm top}, \quad p < p_{\rm eqtop} \\
\Delta T_{\rm bot}\!+\!(\Delta T_{\rm top}\!-\!\Delta T_{\rm bot}) {\frac{\log({\frac{ p}{p_{\rm eqbot}}})}{\log({\frac{p_{\rm eqtop}}{p_{\rm eqbot}}})}}, \\
p_{\rm eqtop}\!\le\!p\!\le\!p_{\rm eqbot} \\
\Delta T_{\rm bot}. \quad p > p_{\rm eqbot}
              \end{array}
              \right.
\end{equation}
Following \cite{zhang-etal-2017}, we set $p_{\rm eqbot}=10$ bar, $p_{\rm eqtop}=10^{-3}$ bar, $\Delta T_{\rm top}$=1000 K, and $\Delta T_{bot}$=0 K.}

There are two ways to distinguish strong and weak external forcings. One is to alter $\Delta T_{\rm eq}$ at the top layer alone as in \cite{komacek-etal-2016}, and the other one is to change ${\rm \tau_{rad}}$ alone. Here, we choose the latter to control the amplitude of external forcing. When the radiative timescale is short, the day-night temperature contrast tends to be large, which represents a strong external forcing. In contrast, when ${\rm \tau_{rad}}$ is greater than relevant dynamical timescales, winds can easily transport heats from the dayside to the nightside, representing a weaker external forcing \citep{komacek-etal-2016}. In brief, changes in the external forcing depend on ${\rm \tau_{rad}}$. Changing ${\rm \tau_{rad}}$, instead of changing $\Delta T_{\rm eq}$, keeps the equilibrium temperature constant in all cases. This method has advantage and limitation: We can fix some atmospheric structures at the reference state, such as the atmospheric stratification, and focus on the other dynamical character. And it is convenient to find the difference between the phase curve depend on actual temperature structure and the phase curve depend on fixed reference temperature structure. The limitation is that different circulation regimes may have different temperature structures, and this makes our simulations and actual observations a bit far away. A few examples of $\tau_{\rm rad}$ used in this study is shown in the right panel of Figure \ref{fig:1}. 

The second forcing is the internal forcing, represented by the thermal perturbations at the deep layers of our model. Following \citet{showman-etal-2019}, we parameterize the interaction between internal convection and the overlying stratified layers as thermal perturbations at isobaric surfaces. We set an extra heat source at the deep model layer to simulate the effects of interactions:

\begin{equation}
\label{eqn50}
{\frac{q_{2}(\lambda,\phi,p,t)}{c_{ p}}}=\left\{
\begin{array}{lcl}
    0, \quad p \leq p_{\rm ftop} \\
    \log(\frac{p}{p_{\rm ftop}})/\log(\frac{p_{\rm bot}}{p_{\rm ftop}})S(\lambda,\phi,t), \quad p > p_{\rm ftop}\\
\end{array}
\right.
\end{equation}
Here $q_{2}$ is heating or cooling from the thermal perturbations, $p_{\rm bot}=$ 400 bars, and $p_{\rm ftop}$ is the pressure of the top level where internal forcing exists (Table \ref{chart:1}). Because altitudes are set at pressure levels, the amplitude of perturbation obeys an exponential decay with the decreasing pressure.

The term $S$ in Equation (\ref{eqn50}) represents a horizontally isotropic, random, and time-evolving thermal perturbation:
\begin{equation}
\label{eqn5}
S(\lambda,\phi,t+\delta t)=rS(\lambda,\phi,t)+\sqrt{1-r^{2}}F(\lambda,\phi,t),
\end{equation}
where
\begin{equation}
\label{eqn6}
F=f_{\rm amp}\sum_{m=1}^{n_{ f}}N_{ n{f}}^{m}(\sin \phi)\cos(m(\lambda+\psi_m)).
\end{equation}
Here, $N_{n}^{m}$ is the normalized associated Legendre polynomials, and Equation (\ref{eqn5}) represents a Markov process \citep{scott-etal-2007,showman-etal-2019}. $f_{\rm amp}$ is a free parameter representing the amplitude of thermal perturbations. $n_f$ is the zonal forcing wavenumber. $\psi_m \in$ $[-180,180]$ is a random number to introduce randomness. In Equation (\ref{eqn5}), the memory coefficient $r$ is defined as:
\begin{equation}
r=1-{\frac{\delta t}{\tau_{\rm d}}},
\end{equation}
where $\tau_{\rm d}$ represents the decaying time of perturbations and $\delta t$ represents the dynamical time step. In this process, $S$ can lose its initial state information over a timescale of $\tau_{\rm d}$. If $\tau_d=\delta t$, $r=0$ and $S$ loses the character of the perturbations in one time step, meaning that $S$ is a white noise and independent of time. On the contrary, if ${\rm \tau_d}$ is large, thermal perturbation patterns will stay for a long time. Thus, the timescale over which global thermal perturbations evolve depends on $\tau_{\rm d}$ for a given dynamical time step $\delta t$.

Key parameters used in this study are listed in Table \ref{chart:1}. Other parameters used in simulations are: specific gas constant $R=3700$ J ${\rm kg^{-1}K^{-1}}$, rotation period of 3.5 Earth days, gravity $g=50$ ${\rm m}$ ${\rm s^{-2}}$, specific heat $c_p=1.3$ $\times$ $10^{4}$ J ${\rm kg^{-1}K^{-1}}$ (represents hydrogen atmosphere) and radius $R_p=9.437$ $\times$ $10^7$ m. The horizontal resolution is C64 in the cubed-sphere grid, which is equivalent to $1.4^\circ \times 1.4^\circ$ in longitude and latitude. Simulations with a higher resolution of C128 ($0.7^\circ \times 0.7^\circ$) are also performed for sensitivity tests (High resolution cases in Table \ref{chart:1}), and the results of the higher resolution simulations are qualitatively similar to those with C64. Thus, we do not show the results of high resolution cases and we will mainly focus on simulations with the C64 resolution. 
We also make some sensitivity tests of $\tau_d$ from $10^4$ to $10^6$ s. The results show that $\tau_d$ has little effect on the atmospheric circulation patterns. And our research focuses on the amplitude of internal and external forcing on the atmospheric circulation, so the character of internal forcing, such as $\tau_d$, is not in the scope of this research, and we do not show the results.
We choose $n_f=20$ and $\delta n=1$ for all simulations, which is because we want to excite a global perturbation and we don't have much constraint on them.
The maximum pressure $p_{\rm bot}$ is 400 bars, and the minimum pressure $p_{\rm top}$ is $10^{-4}$ bar in our model, with 80 levels in the log-pressure space. The dynamical time step ${\rm \delta t}$ is 20 seconds for all simulations. The decay timescale ${\rm \tau_{d}}$ is ${\rm 10^5}$ s for all simulations. Our model is a dry air, cloud-free model, without effects of condensation and chemical reactions. The amplitude of thermal perturbations $f_{\rm amp}$ has a directly relationship with the interior heat fluxes. According to the theory of mixing lengths and scaling analysis \citep{showman-etal-2019}, for the case with internal heat flux ${\rm \sim 10}$ ${\rm W}$ ${\rm m^{-2}}$, we obtain the perturbation amplitude of $f_{\rm amp}=$ ${\rm 10^{-6}}$ ${\rm K}$ ${\rm s^{-1}}$, and for the case with internal heat flux ${\rm 10^3-10^4}$ ${\rm W}$ ${\rm m^{-2}}$, we obtain the amplitude of ${\rm 10^{-4}}$ ${\rm K}$ ${\rm s^{-1}}$.

\begin{table*}[!htb]
\centering
\scriptsize
\caption{ Parameters used in simulations. Here, the abbreviates are: In: internal forcing amplitude; Ex: day-night (external) stellar forcing; H: high resolution; Shallow: shallower boundary layer; med: medium; $p_{\rm ftop}$: the pressure of the top level where internal forcing exists; $\tau_{\rm radtop}$: radiative time scale of top layer; $f_{\rm amp}$: thermal perturbation amplitude.
For example, ${\rm In_{0}Ex_{strong}}$ means zero internal forcing and strong external forcing}
\setlength{\tabcolsep}{1mm}{
\begin{tabular}{|ccccc|}
\hline
Case                                     & $f_{\rm amp}$ (${\rm K}$ ${\rm s^{-1}}$) & $\tau_{\rm radtop}$ (s) & $p_{\rm ftop}$ (bars) & Horizontal resolution \\ \hline
\multicolumn{5}{|c|}{Cases of different external/internal forcing}                                                                                          \\ \hline
${\rm In_{0}Ex_{strong}}$                      & 0                & $10^5$                 & 200                                      & C64           \\
${\rm In_{0}Ex_{med}}$                         & 0                & $10^6$                 & 200                                      & C64           \\
${\rm In_{0}Ex_{weak}}$                        & 0                & $10^7$                 & 200                                     & C64           \\
${\rm In_{weak}Ex_{strong}}$                   & $10^{-6}$                & $10^5$         & 200                                      & C64           \\
${\rm In_{weak}Ex_{med}}$                      & $10^{-6}$                & $10^6$         & 200                                     & C64           \\
${\rm In_{weak}Ex_{weak}}$                     & $10^{-6}$                & $10^7$         & 200                                     & C64           \\
${\rm In_{med}Ex_{strong}}$                    & $10^{-5}$                & $10^5$         & 200                                     & C64           \\
${\rm In_{med}Ex_{med}}$                       & $10^{-5}$                & $10^6$         & 200                                     & C64           \\
${\rm In_{med}Ex_{weak}}$                      & $10^{-5}$                & $10^7$         & 200                                      & C64           \\
${\rm In_{strong}Ex_{strong}}$                 & $10^{-4}$                & $10^5$         & 200                                      & C64           \\
${\rm In_{strong}Ex_{med}}$                    & $10^{-4}$                & $10^6$         & 200                                      & C64           \\
${\rm In_{strong}Ex_{weak}}$                   & $10^{-4}$                & $10^7$         & 200                                      & C64           \\ \hline
\multicolumn{5}{|c|}{High resolution cases}                                                                                                                 \\ \hline
${\rm In_{strong}Ex_{strong}H}$                & $10^{-4}$                & $10^5$         & 200                                     & \textbf{C128}           \\
${\rm In_{strong}Ex_{med}H}$                   & $10^{-4}$                & $10^6$         & 200                                    & \textbf{C128}           \\
${\rm In_{strong}Ex_{weak}H}$                  & $10^{-4}$                & $10^7$         & 200                                    & \textbf{C128}           \\ \hline
\multicolumn{5}{|c|}{Changing the level of bottom forcing}                                                                                                  \\ \hline
$\rm {In_{med}Ex_{strong}Shallow}$             & $10^{-5}$                & $10^5$         & \textbf{100}                                      & C64           \\
$\rm {In_{med}Ex_{med}Shallow}$                & $10^{-5}$                & $10^6$         & \textbf{100}                                      & C64           \\
$\rm {In_{med}Ex_{weak}Shallow}$               & $10^{-5}$                & $10^7$         & \textbf{100}                                     & C64           \\ \hline
\end{tabular}}
\label{chart:1}
\end{table*}

\begin{figure*}[t]
\centering
\includegraphics[width=0.95\textwidth]{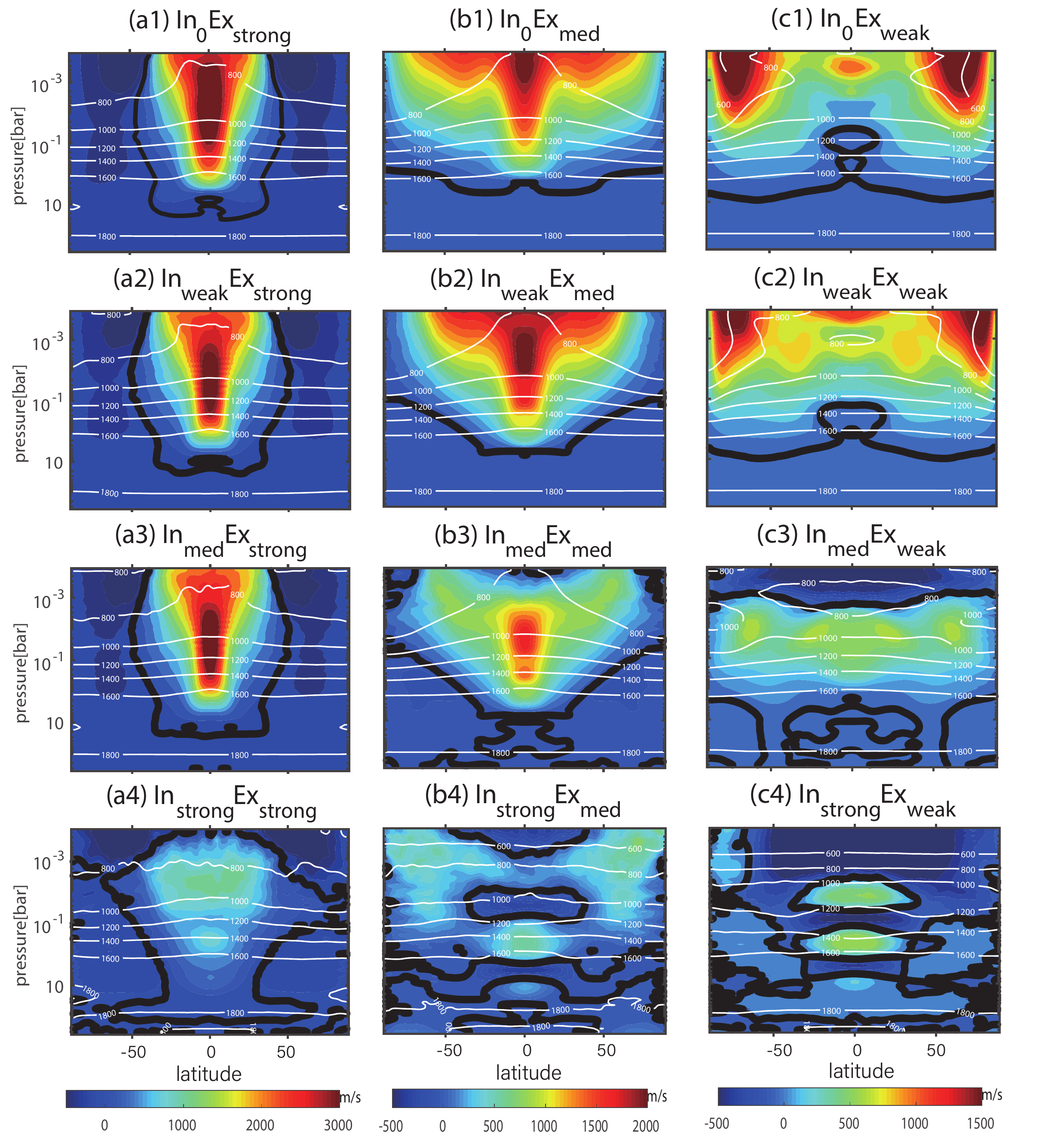}\\
\caption{ Vertical cross-sections of zonal-mean zonal winds for various external and internal forcings. Left column: strong external forcing (strong day-night irradiation contrast) with short radiative timescales at the top of atmosphere ($\tau_{\rm radtop}=10^5$ ${\rm s}$, a1-a4); middle column: medium external forcing ($\tau_{\rm radtop}=10^6$ ${\rm s}$, b1-b4); right column: weak external forcing ($\tau_{\rm radtop}=10^7$ ${\rm s}$, c1-c4). The rows from top to bottom are for different internal thermal perturbation from weak to strong ($f_{\rm amp}=0, 10^{-6}, 10^{-5}$, and $10^{-4}$ ${\rm K}$ ${\rm s^{-1}}$). White lines show zonal-mean temperature contours, and thick solid black lines show zero velocity contours. Color bar unit is ${\rm m}$ ${\rm s^{-1}}$, and the unit of white lines is ${\rm K}$. All results are snapshots at day 3500 in simulations.}
\label{fig:2}
\end{figure*}

\begin{figure*}
\centering
\includegraphics[width=0.93\textwidth]{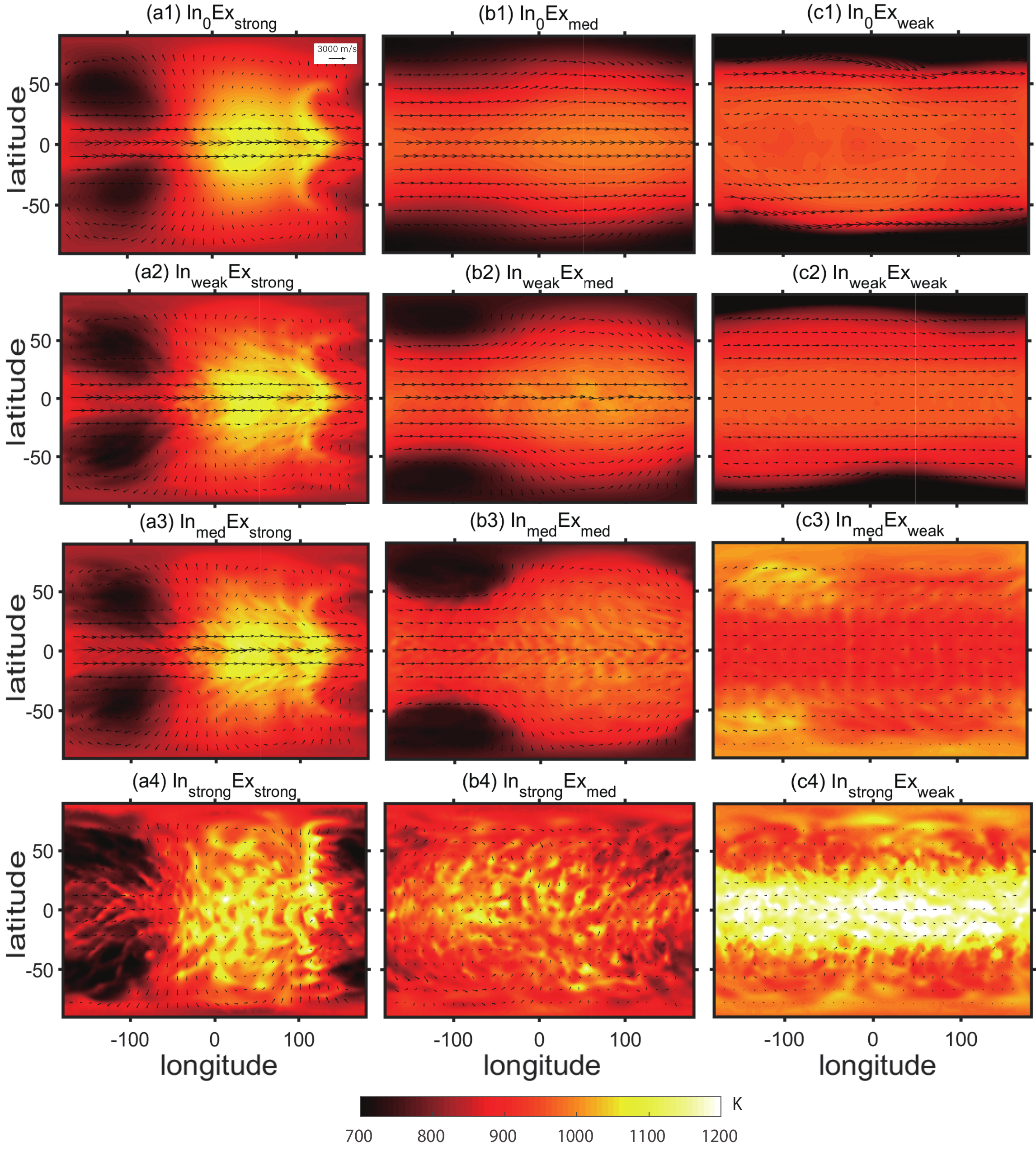}\\
\caption{Same as Figure \ref{fig:2}, except for temperatures (color shading) and winds (arrows) at 10 mbar. Results are snapshots at 3500 model days. Color bar unit is K.}
\label{fig:3}
\end{figure*}

The relative roles of stationary and transient eddies in the momentum balance of zonal-mean zonal flows are diagnosed by the following equation \citep[e.g.][]{andrews-1987}:
\begin{equation}
\begin{split}
{\frac{\partial \bar{u}}{\partial t}}=-X+f\bar{v}-{\frac{\bar{v}}{a \cos\phi}}{\frac{\partial}{\partial\phi}}(\bar{u}\cos\phi)-\bar{\omega}{\frac{\partial\bar{u}}{\partial p}}\\
-{\frac{1}{a \cos^2{\phi}}}{\frac{\partial}{\partial\phi}}(\overline{u^*v^*}\cos^2\phi)-{\frac{\partial}{\partial p}}(\overline{u^*\omega^*})\\
-{\frac{1}{a \cos^2\phi}}{\frac{\partial}{\partial\phi}}(\overline{u'v'}\cos^2\phi)-{\frac{\partial}{\partial p}}(\overline{u'\omega'}),
\end{split}
\label{eqn7}
\end{equation}
where $\bar{u},\bar{v},\bar{\omega}$ represent time- and zonal-mean zonal velocity, meridional velocity, and vertical velocity, respectively. $X$ denotes friction. The second line of Equation (\ref{eqn7}), the term of $\overline{u^*v^*}$, represents the acceleration due to horizontal momentum convergence of stationary eddies, and the term of $\overline{u^*\omega^*}$ represents the acceleration due to vertical momentum convergence of stationary eddies. The third line includes accelerations of horizontal and vertical momentum convergences of transient eddies. Total eddies are obtained by subtracting the zonal-time-mean flow from the velocity fields, and further taking a time-averaging of eddy fields to obtain stationary eddies. Finally, transient eddies are obtained by subtracting stationary eddies from total eddies. 
We think that most of the stationary eddies are related to external day-night forcing, and the transient eddies are related to the internal bottom forcing, but it does not rule out that some transient eddies are intrinsically excited by the circulation.

\section{RESULTS}
\label{Result}

The key results of our simulations are shown in Figures \ref{fig:2} and \ref{fig:3}. Figure \ref{fig:2} shows zonal-mean zonal winds, and Figure \ref{fig:3} shows global distributions of temperatures at 10 mbar. From left to right, the radiative timescales vary from $10^5$ s to $10^7$ s, representing external forcing from strong to weak. From top to bottom, internal thermal perturbation amplitudes vary from 0 ${\rm K}$ ${\rm s^{-1}}$ (top row, a1-c1) to $10^{-4}$ ${\rm K}$ ${\rm s^{-1}}$ (bottom row, a4-c4), representing internal forcing from weak to strong. The results in Figures \ref{fig:2} and \ref{fig:3} fall into three regimes: The left column (a1, a2, a3, a4) belongs to the equatorial superrotation regime. Plots c1 and c2 belong to the midlatitude-jet regime. { Plots b1, b2 and b3 belong to the transition from the superrotation to the midlatitude-jet regime}, and plots b4, c3, and c4 belong to the regime of quasi-periodic equatorial oscillation. We shall address the three regimes in detail in the following. Notice that, Plots a4, b4 and c4 show stochastic temperature fluctuations in Figure \ref{fig:3} even they are not from the perturbation layers. Most of those fluctuations are from upward-propagating inertia gravity waves, because the small scale waves deviate significant from the geostrophic balance.

\subsection{Superrotation regime (Cases ${\rm In_{0}Ex_{strong}-In_{strong}Ex_{strong}}$)}
\label{1 regime}

The cases with strong external forcing, regardless of the strength of internal forcing, all exhibit equatorial eastward jets. The jet is strong when the internal forcing amplitude is ${\rm \leq 10^{-5}}$ ${\rm K}$ ${\rm s^{-1}}$, but becomes significantly weaker when the internal forcing increases to ${\rm 10^{-4}}$ ${\rm K}$ ${\rm s^{-1}}$. In the following, we explain the strong jet cases (Figures \ref{fig:2}a1-a3) first, and then the weak jet case (Figures \ref{fig:2}a4).

\subsubsection{Strong superrotation}
\label{1.1 regime}

The maximum zonal-mean wind speed in Figures \ref{fig:2}a1-a3 is up to 3000 ${\rm  m}$ ${\rm s^{-1}}$. The jet core in these cases moves to slightly higher pressures when the internal thermal forcing becomes stronger. The overall atmospheric circulation pattern remains steady {even with time-varying bottom perturbations}. This is the typical pattern of atmospheric circulations of hot Jupiters, as shown in previous studies \citep{showman2020review}. In these cases, the external forcing is dominant. It excites stationary Kelvin and Rossby waves, resulting in eastward shift of the hot spot at the equator and westward shift of the cold spot at midlatitudes (Figures \ref{fig:3}a1-a3). The resulting stationary Rossby-Kelvin wave pattern induces the northwest-southeast tilting of horizontal velocities in the northern hemisphere and northeast-southwest tilting in the southern hemisphere. Such an eddy-wind tilting indicate southward and northward eddy momentum transports in the North and South hemispheres, respectively. These eddy-momentum transports generate eastward acceleration on equatorial winds, which maintains the equatorial superrotation \citep{showman-etal-2011, hu-etal-2013,tsai-etal-2014, hammond-etal-2018, debras-etal-2020,wang-yang-2020}. Even for moderate thermal perturbations, i.e., case ${\rm In_{med}Ex_{strong}}$ (Figure \ref{fig:3}a3), the typical stationary Rossby-Kelvin wave pattern remains, except for some small-scale disturbances.

\subsubsection{Weak superrotation}

\begin{figure*}
\centering
\includegraphics[width=0.85\textwidth]{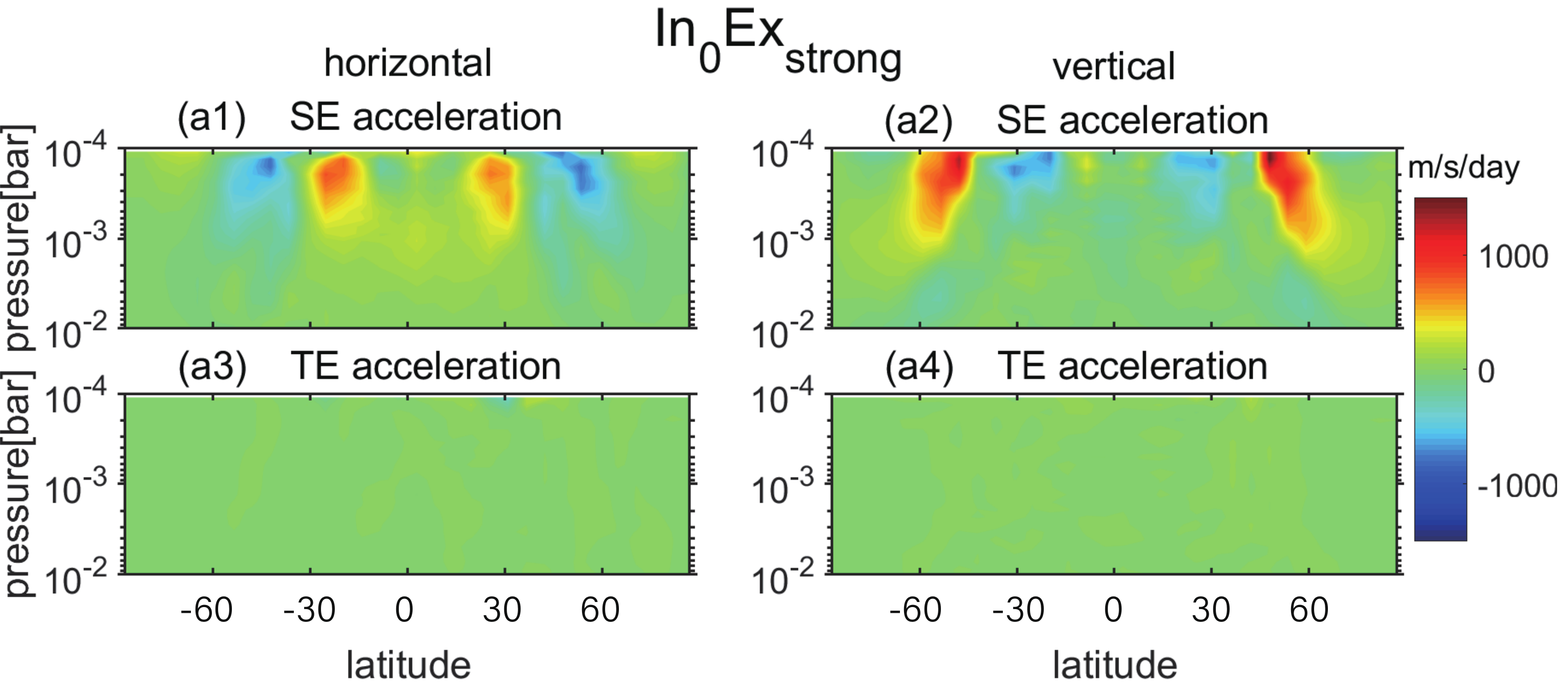}\\
\includegraphics[width=0.85\textwidth]{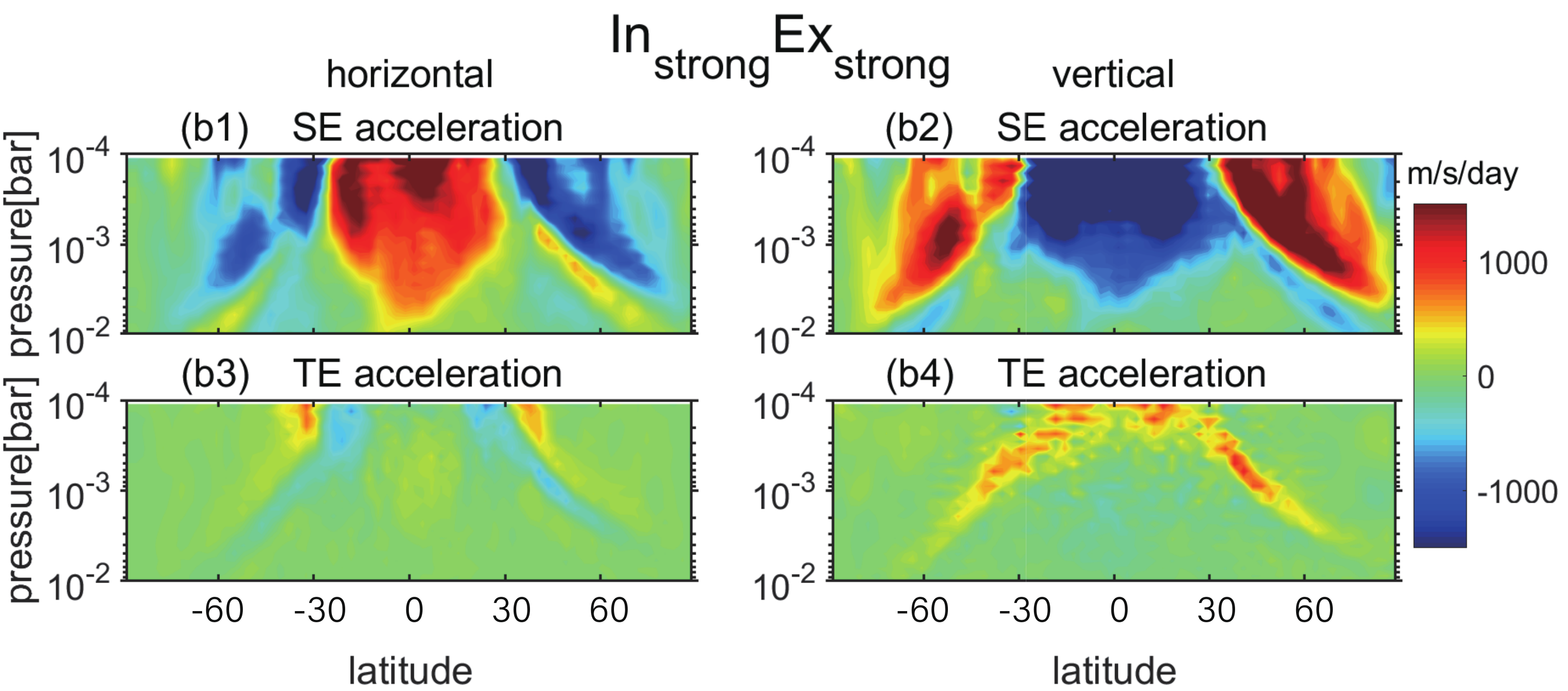}
\caption{ Divergences of stationary and transient eddy terms in Equation (\ref{eqn7}) for cases ${\rm In_0Ex_{strong}}$ and ${\rm In_{strong}Ex_{strong}}$. SE denotes Stationary eddies, and TE denotes Transient eddies. The duration of time-averaging of eddy fields is 1 month to obtain stationary eddies. Horizontal acceleration represents $-{\frac{1}{acos^2{\phi}}}{\frac{\partial}{\partial\phi}}(\overline{u'v'}cos^2\phi)$ and vertical acceleration represents $-{\frac{\partial}{\partial p}}(\overline{u'\omega'})$. Color bar unit is ${\rm m}$ ${\rm s^{-1}}$ ${\rm day^{-1}}$.}
\label{fig:acc}
\end{figure*}

\begin{figure*}
\centering
\includegraphics[height=7.0cm,width=14.5cm]{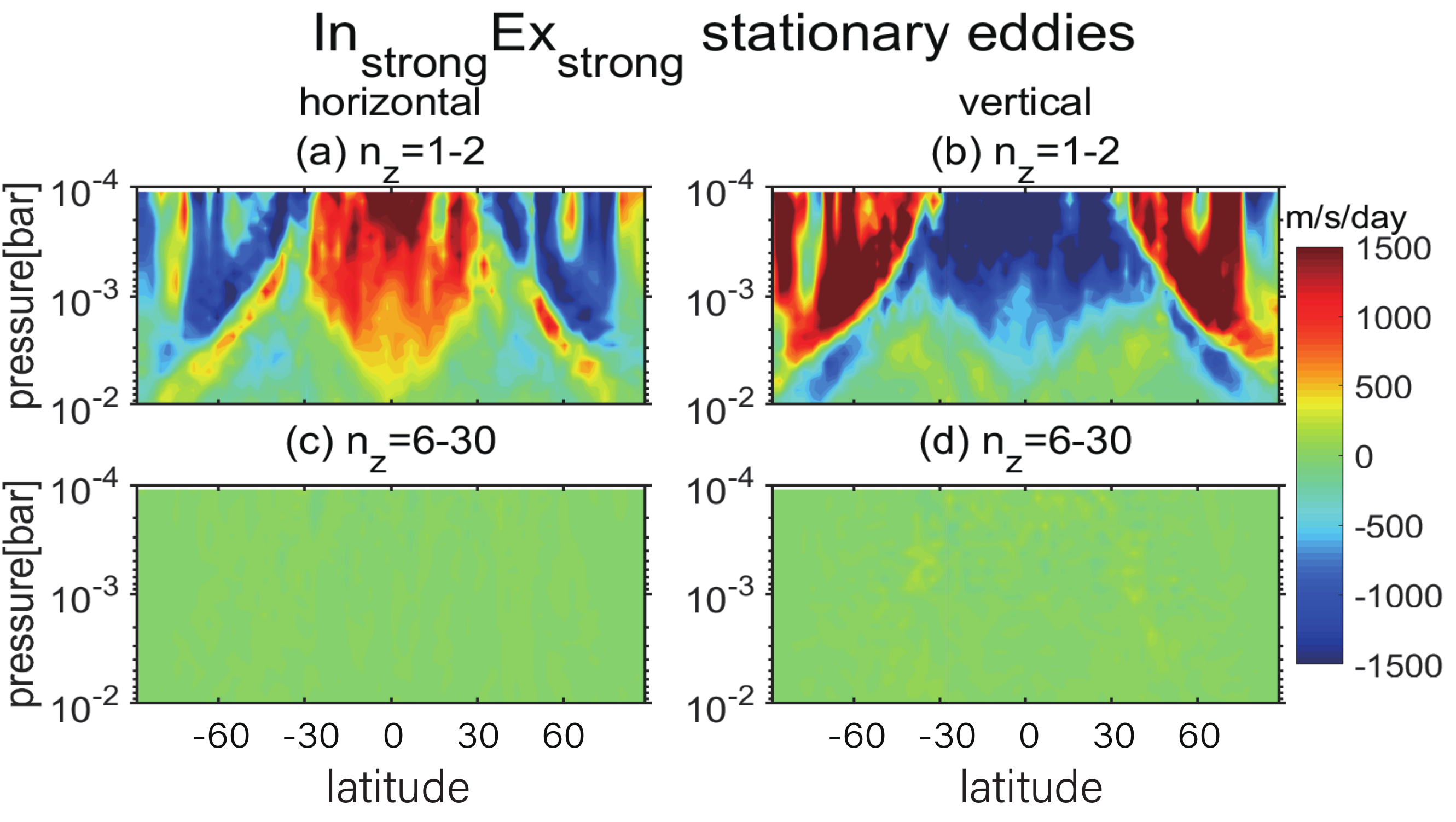}\\
\includegraphics[height=7.0cm,width=14.5cm]{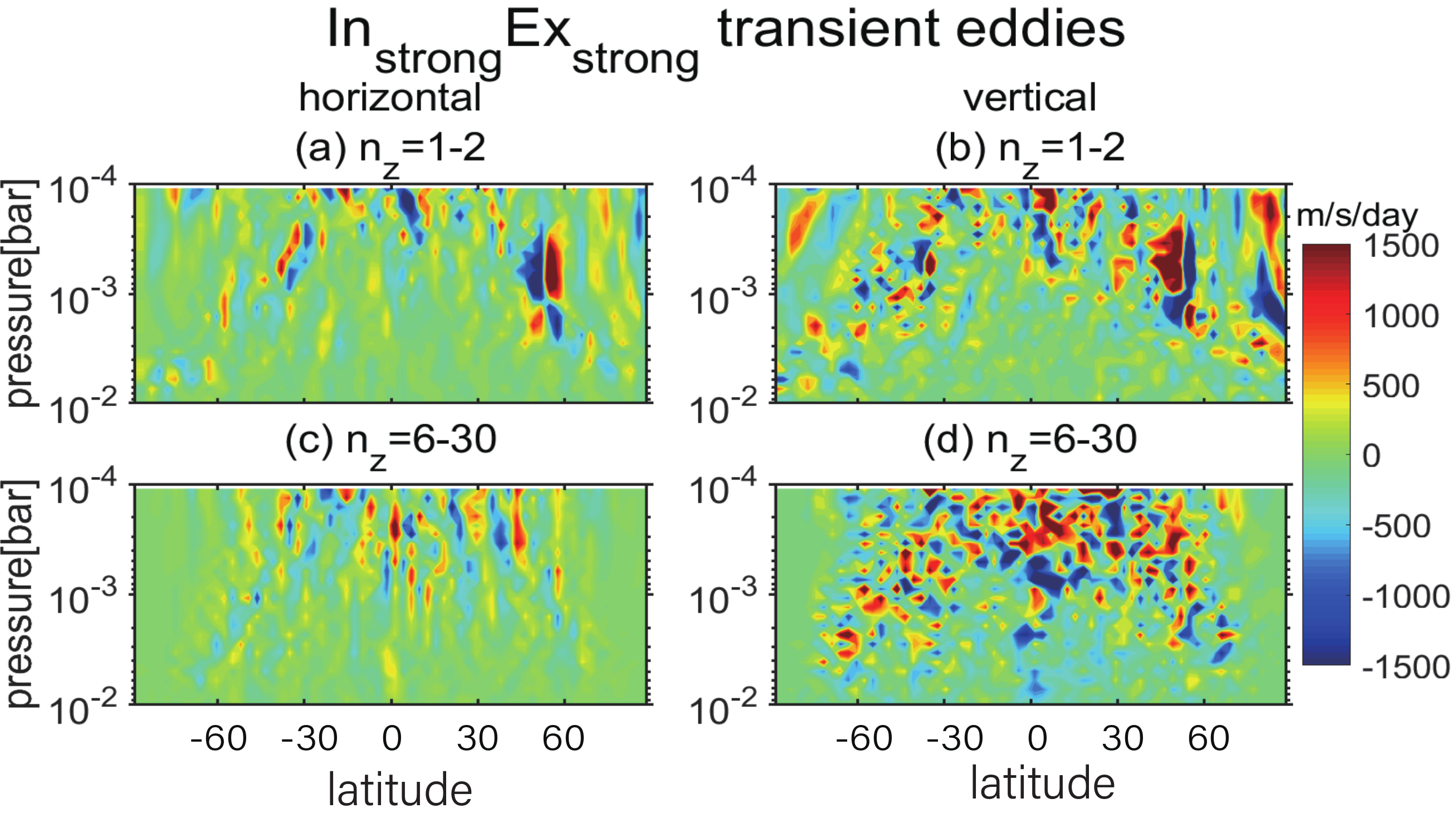}
\caption{ Stationary and transient eddy-acceleration of long (with zonal wavenumber $n_{z}=$ 1, 2) and short (with zonal wavenumber $n_{z}=$ 6-30) waves with Fourier analysis. Horizontal acceleration is $-{\frac{1}{acos^2{\phi}}}{\frac{\partial}{\partial\phi}}(\overline{u(n_{z})'v(n_{z})'}cos^2\phi)$ and vertical acceleration is $-{\frac{\partial}{\partial p}}(\overline{u(n_{z})'\omega(n_{z})'})$. Color bar unit is ${\rm m}$ ${\rm s^{-1}}$ ${\rm day^{-1}}$.}
\label{fig:acc1}
\end{figure*}

The equatorial eastward jet remains in case ${\rm In_{strong}Ex_{strong}}$ (Figure \ref{fig:2}a4). However, the maximum zonal-mean wind speed drops to $\sim$ 1000 ${\rm m}$ ${\rm s^{-1}}$. It is indicative of the role of the strong internal forcing in decelerating the equatorial eastward jets.

To study how the strong internal forcing alters the equatorial jet, we compare the acceleration terms in Equation (\ref{eqn7}) for strong and weak external forcings, which are shown in Figure \ref{fig:acc}. In the case of ${\rm In_{0}Ex_{strong}}$ (Figure \ref{fig:acc}a1), stationary eddies generate eastward acceleration to the equatorial jet, by transporting horizontal eddy momentum fluxes, and the influences of transient eddies can be negligible.

In the case of ${\rm In_{strong}Ex_{strong}}$ (Figure \ref{fig:acc}b), horizontal stationary eddy momentum fluxes have large contributions to the eastward acceleration, while vertical stationary eddy momentum fluxes cause strong deceleration at the upper tropical levels. The transient eddies cause slight eastward acceleration by horizontal eddy momentum transports and eastward accelerations (${\rm \sim 5 \times 10^2}$ ${\rm m}$ ${\rm s^{-1} day^{-1}}$) by vertical momentum transports, especially at upper levels. The eastward acceleration of transient eddies may be caused by the damping of Kelvin waves that are generated by internal forcing. The eastward equatorial jet is transparent to Rossby-gravity waves. Therefore, there is no westward acceleration driven by transient eddies.

In Figure \ref{fig:acc}, the accelerations are only shown at low pressures because the temperature difference is small due to the longer $\tau_{\rm rad}$ at 10 bars, and then the eddies near 10 bars are weaker than the eddies at low pressure. The zonal flow is nearly time-steady. If we include all the acceleration, there is a nearly balance of all the forces. The numerical dissipation (the Shapiro filter) has very minor contribution to the angular momentum budget as well.

We apply the Fourier analysis to zonal winds in order to separate eddies with different zonal length scales in the case of ${\rm In_{strong}Ex_{strong}}$, and then separate the wave components into stationary and transient eddies (Figure \ref{fig:acc1}). \citet{mendonca-2020} mentioned that eddies with zonal wavenumber $1$ and $2$ likely correspond to the external forcing of diurnal and semidiurnal tides. However, we find that, transient eddies brings considerable components to the $n_z=1-2$ case, although stationary eddies have a decisive effect on the the $n_z=1-2$ case. Eddies with large zonal wavenumbers (e.g., $6-30$) are totally corresponding to the transient eddies. It can be seen in Figure \ref{fig:acc1} that: 1. Transient eddies contribute much to $n_z=1,2$ case, considering transient eddies may be caused by thermal perturbation,  the eddies with zonal wavenumber $1, 2$ could correspond to the internal forcing, rather than all those eddies correspond to external forcing; 2. Most of westward accelerations to the equatorial jet are caused by vertical momentum fluxes of large stationary eddies. Horizontal momentum fluxes of large stationary eddies play a crucial role in the establishment and maintaining of the equatorial jet. Briefly, externally forced waves dominate the weak equatorial superrotation in Figure \ref{fig:2}d. Although internally forced waves have effects on equatorial jets, they cannot dominate the behavior of equatorial jets. Stationary eddies generate eastward (westward) acceleration by horizontal (vertical) momentum transports.

It is out of our initial expectation that transient eddies would directly contribute to westward acceleration of the equatorial flow, slowing down the jet. The diagnose described above seems to be more subtle than our expectation. Although the internal forcing itself does not show strong influences on this pattern by the upward propagating equatorial waves, the internal forcing may increase the strength of stationary eddies, strongly impacting on diurnal and semidiurnal waves driven by strong external forcing. The net effect is to cause the formation of the weak equatorial superrotation. It is a result of the offset of two large acceleration terms with almost equal magnitudes. Thus, the exact effect of the internal forcing is difficult to diagnose from the fully nonlinear simulations. An additional effect of the largely weakened equatorial eastward jet is that it transports less heats to the nightside, causing a stronger day-night temperature contrast. This would have implications for phase-curve observations, which will be further discussed in Section \ref{Lightcurve}.

\subsection{Midlatitude jet regime (Cases ${\rm In_{0}Ex_{weak}}$ and ${\rm In_{weak}Ex_{weak}}$)}
\label{2 regime}

For the cases with both weak external and internal forcings, the equatorial eastward jet becomes weaker, while strong eastward jets form at middle and high latitudes (Figures \ref{fig:2}c1-c3). In these cases, the radiative timescale is longer than the advection timescale, so that heat is more readily transported from the dayside to the nightside, causing weaker day-night temperature contrasts, compared to the cases of strong external forcing. This caused weaker stationary waves and thus weakened equatorward transports of eddy momentum fluxes. As a result, the equatorial eastward jet is largely weakened. 

\begin{figure}
\includegraphics[width=0.5\textwidth]{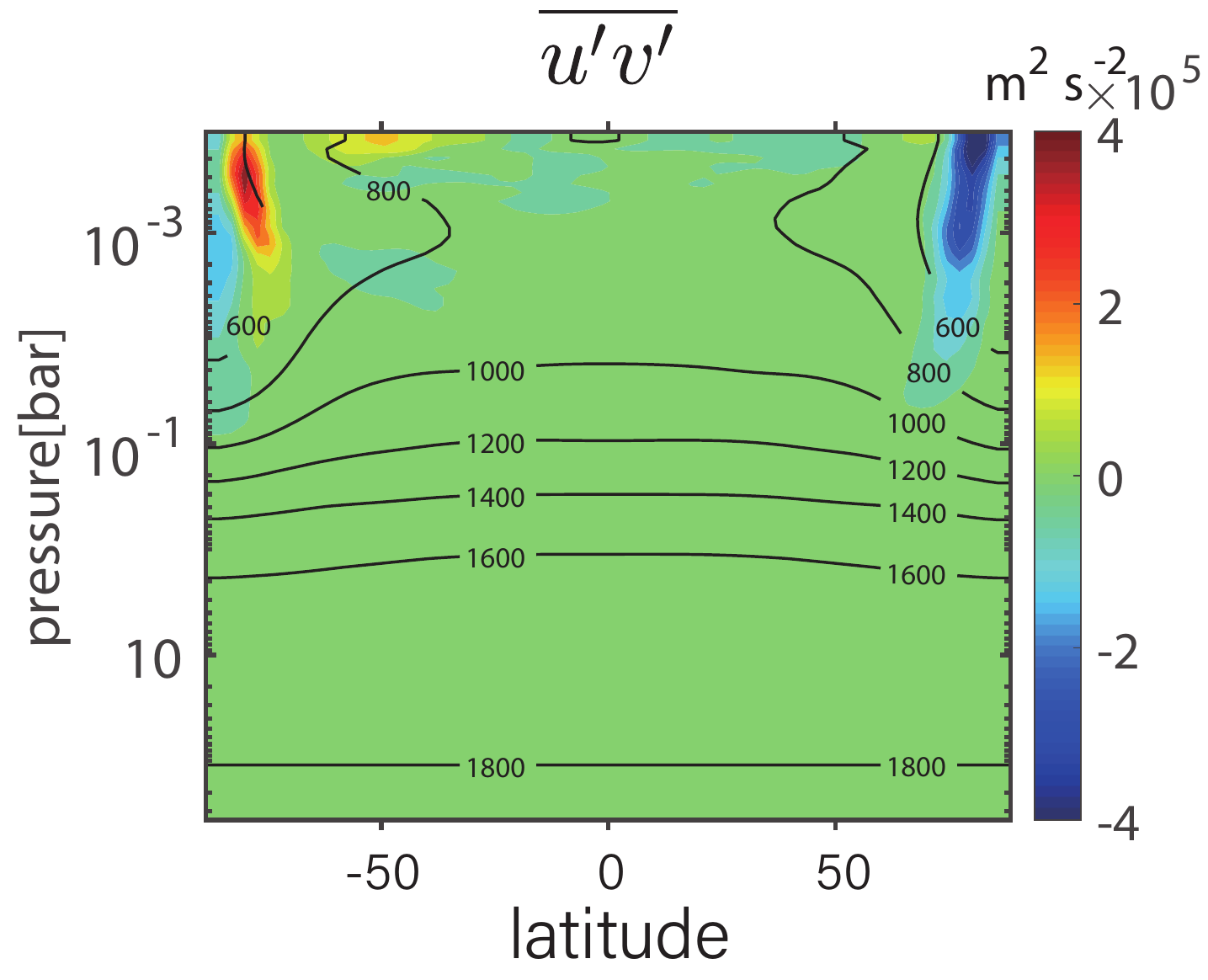}
\caption{ The horizontal momentum fluxes by the transient eddies ($\overline{u'v'}$) of case ${\rm In_{0}Ex_{weak}}$ (Figure \ref{fig:2}c1). The black solid lines show the zonal mean temperature.}
\label{fig:epoleacc}
\end{figure}

The circulation pattern of the midlatitude jet regime is qualitatively similar to that in previous studies that have similarly long radiative timescales \citep{perez-etal-2013, komacek-etal-2016,tan-etal-2020}. The strong eastward jets at middle and high latitudes could be caused due to two major reasons. First, for the cases with weak external forcing, the role of equator-to-pole temperature contrasts becomes more significant compared with that with strong external forcing. Especially, the largest latitudinal temperature gradient is around $\pm 60^{\circ}$ in latitude, which causes the eastward jets due to the thermal wind balance \citep{holton-2004}. Second, the jets are also driven by baroclinic eddies that are caused by meridional temperature gradients. The transient baroclinic eddies generate  momentum flux convergence at high latitudes, which accelerates the middle and high-latitude eastward jets \citep[e.g.][]{charney-1947,farrell-etal-2008}. We make a diagnose of the horizontal momentum fluxes by the transient eddies of case ${\rm In_0Ex_{weak}}$ in Figure \ref{fig:epoleacc} and our diagnose analysis does show the equatorward momentum fluxes generated by the transient eddies at $\pm 70^\circ$, $\sim 10^{-3}$ bar.

Figures \ref{fig:2}b1-b3 show the transitional states from the superrotation regime to the midlatitude-jet regime, which are associated with medium external forcing and weak internal forcing. The equatorial eastward jet becomes weaker and moves to lower levels with increasing internal thermal forcing. At the same time, mid-latitude eastward jets emerge. The maximum equatorward momentum fluxes are at ${\pm 60^\circ}$ in latitude for cases of ${\rm In_{0}Ex_{med}}$ and ${\rm In_{weak}Ex_{med}}$, in contrast to that at ${\pm 30^\circ}$ in latitude for cases with strong external forcing (${\rm In_{0}Ex_{strong}}$ and ${\rm In_{weak}Ex_{strong}}$). 

\subsection{Regime of quasi-periodic oscillations (Cases ${\rm In_{strong}Ex_{med}}$, ${\rm In_{med}Ex_{weak}}$ and ${\rm In_{strong}Ex_{weak}}$)}
\label{3 regime}

\begin{figure}
\centering
\includegraphics[width=0.5\textwidth]{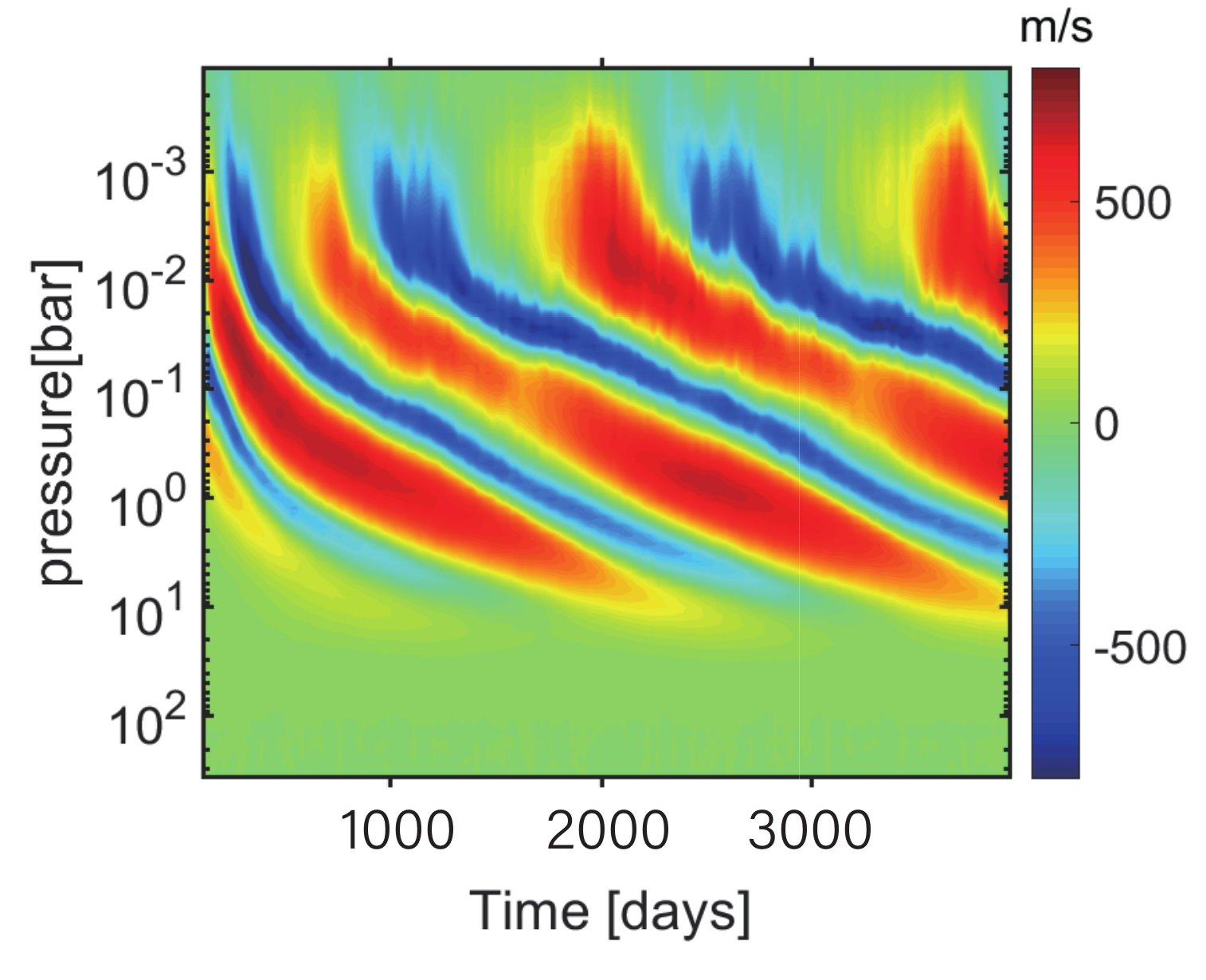}\\
\caption{Vertical-time cross-section of equatorial zonal-mean zonal winds (Earth days) in case ${\rm In_{strong}Ex_{weak}}$. Blue colors denote westward winds (negative values), and red-yellow colors denote eastward winds (positive values).}
\label{fig:qbotime1}
\end{figure}

We have seen in Figure \ref{fig:2}b4, c3, and c4 that strong and vertically coherent equatorial eastward jets no longer exist for the cases of ${\rm In_{strong}Ex_{med}}$, ${\rm In_{med}Ex_{med}}$, and ${\rm In_{strong}Ex_{weak}}$, and that the eastward equatorial jets are largely weakened. Our further analysis reveals a quasi-periodic reversal of eastward and westward equatorial winds (i.e., quasi-periodic oscillations). This phenomenon is illustrated in Figure \ref{fig:qbotime1}, which shows the time-pressure cross-section of equatorial zonal-mean zonal winds for the case of ${\rm In_{strong}Ex_{weak}}$. It takes about 3000 Earth days for eastward and westward winds to propagate from the top to about $10$ bars. the migrating jet is dissipated at pressure higher than 10 bars and new jet are generated near the top layer. The maximum wind speed is about $600$ ${\rm m}$ ${\rm s^{-1}}$. There are three quasi-periodic reversals of eastward and westward winds over the 4000 Earth days of simulations. Each period of the wind reversal is about 1500 days, about 4 Earth years. Such a phenomenon is similar to the Quasi-Biennial Oscillation (QBO) in the middle and lower equatorial stratosphere of the Earth atmosphere, which has the quasi-period of about 28 Earth months \citep{baldwin-etal-2001, yang-yu-2016}. It is noticed that the westward wind lasts nearly twice longer than the eastward wind.

The QBO phenomenon in Earth's atmosphere is very well studied \citep{lindzen-etal-1968, lindzen-etal-1970, lindzen-etal-1971, baldwin-etal-2001}. It is caused by interactions between the tropical zonal flow and tropical waves, such as Kelvin waves, mixed Rossby-gravity waves, and inertial-gravity waves. 
Equatorial Kelvin waves are special boundary waves trapped near the equator, and the Rossby-gravity waves are special solution of the Rossby waves near the equator. When the phase velocity is negative, the Rossby-gravity waves behave like the Rossby wave, and when the phase velocity is positive, they are similar to the inertial gravity waves. 
Those waves are generated in the troposphere and vertically propagate into the stratosphere. 
Kelvin waves have eastward phase speed with eastward momentum fluxes ($u'w'>0$). Rossby-gravity waves also show $u'w'>0$, but they have strong poleward heat transports, and the net effect is to transport the westward momentum fluxes to jets.
For the eastward QBO phase, eastward propagating waves break up at the critical levels of the eastward equatorial zonal flow, generating eastward acceleration for the zonal flow and leading to downward migration of eastward winds. On the other hand, the eastward zonal flow is transparent to westward propagating waves. So that, westward propagating waves can reach the upper stratosphere and damp there, causing downward migration of the westward winds. 

\begin{figure*}
\centering
\includegraphics[width=0.9\textwidth]{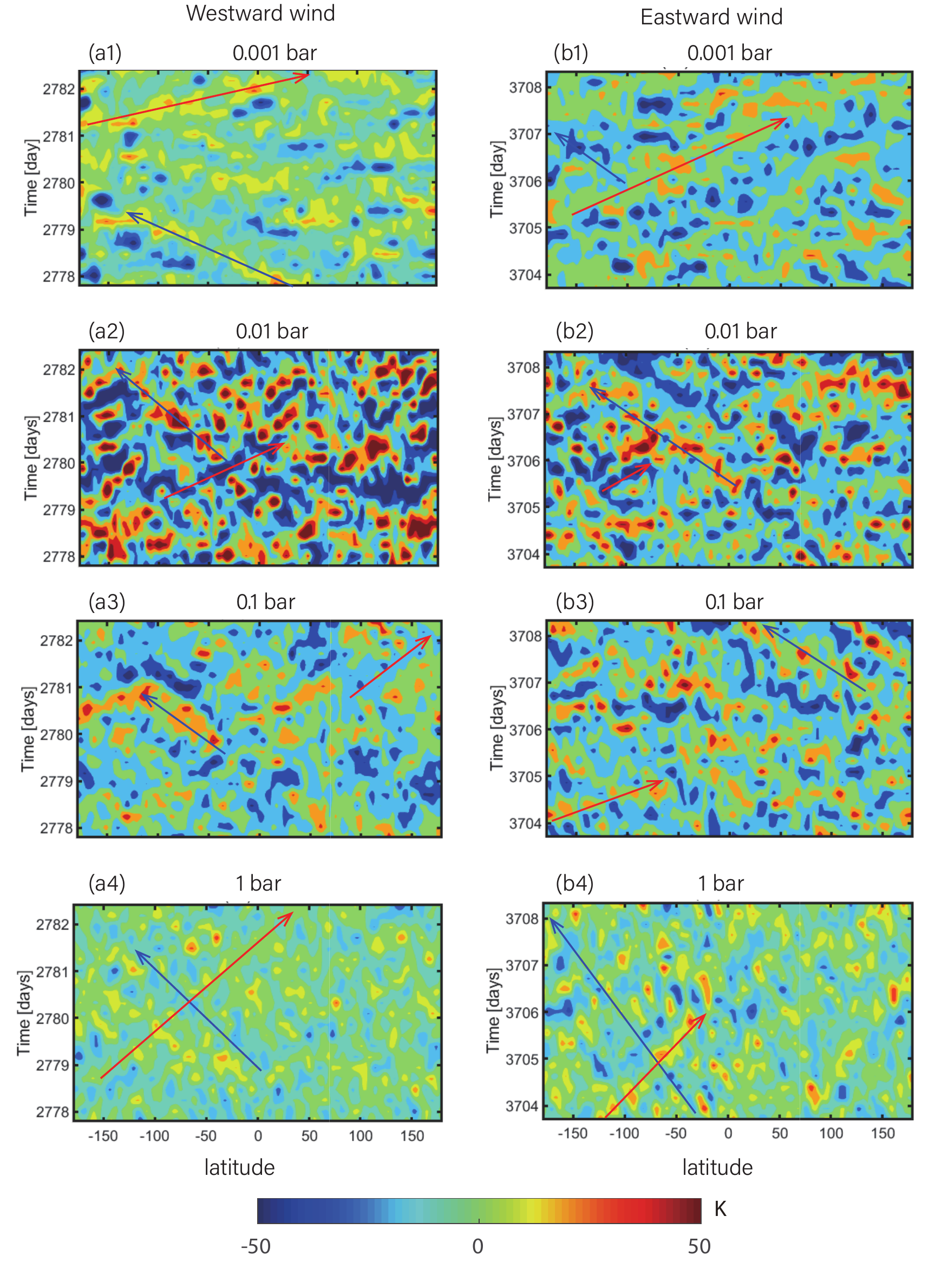}
\caption{ Time evolution of equatorial temperature anomalies at different pressure levels for the case of ${\rm In_{strong}Ex_{weak}}$. Left panels: westward wind phase (a1-a4), and right panels: eastward wind phase (b1-b4). These snapshots show eddies at pressure levels between $10^{-3}$ bar and 1 bar. Red arrows indicate propagation directions of eastward waves, and blue arrows indicates propagation directions of westward waves. Color bar unit is K.}
\label{fig:6}
\end{figure*}

Figure \ref{fig:6} shows time evolution of equatorial temperature anomalies at different pressure levels for westward and eastward winds, respectively. Note that the Kelvin waves propagate faster than the Rossby waves and long Rossby-gravity waves. Thus, the slope of the topleft-downright tilting is less than the slope of the topright-downleft tilting. Figures \ref{fig:6}a1-a4(b1-b4) show that the westward (eastward) waves, with slow phase speeds, damp in the westward (eastward) jets. { It appears that zonal winds act as a filters for slow wave}. When zonal winds are in the westward (eastward) phase at lower pressure levels, the westward (eastward) propagating waves become significantly faster at $10^{-3}$ bar, while the eastward (westward) propagating waves do not change their phase speeds. The slower waves encounter the critical levels of the jets, and the fast-propagating waves are relatively unaffected by the jets. 

Figure \ref{fig:6} also shows that wave amplitudes grow from about $20$ ${\rm K}$ at $10^{-1}$ bar to about $100$ ${\rm K}$ at $10^{-2}$ bar, and then decrease to about $20-40$ ${\rm K}$ at the top. The wave amplitudes are constrained by the mean flows \citep{chao-etal-1984,fritts-1984}. For a monochromatic wave, its eddy velocity $u'$ is supposed to meet the condition between wave phase speed $c$ and the mean-flow speed $U$:
\begin{equation}
\left| u' \right| \leq \left| c-U \right|
\label{condition}
\end{equation}
If this condition is not satisfied, the monochromatic wave would release wave energy. In addition, wave amplitudes increase with altitude because of the decrease of the background air density \citep{vanzandt-etal-1989}. The net effect is that amplitudes of most waves with moderate phase speeds, increase below the jet core, and decrease after passing through the jet core. In brief, Figure \ref{fig:6} shows that the maximum absorption of equatorial waves is between $10^{-2}$ and $10^{-3}$ bar. The maximum wave absorption causes the maximum eastward or westward acceleration.

\begin{figure*}
\centering
\includegraphics[height=10.0cm,width=17.0cm]{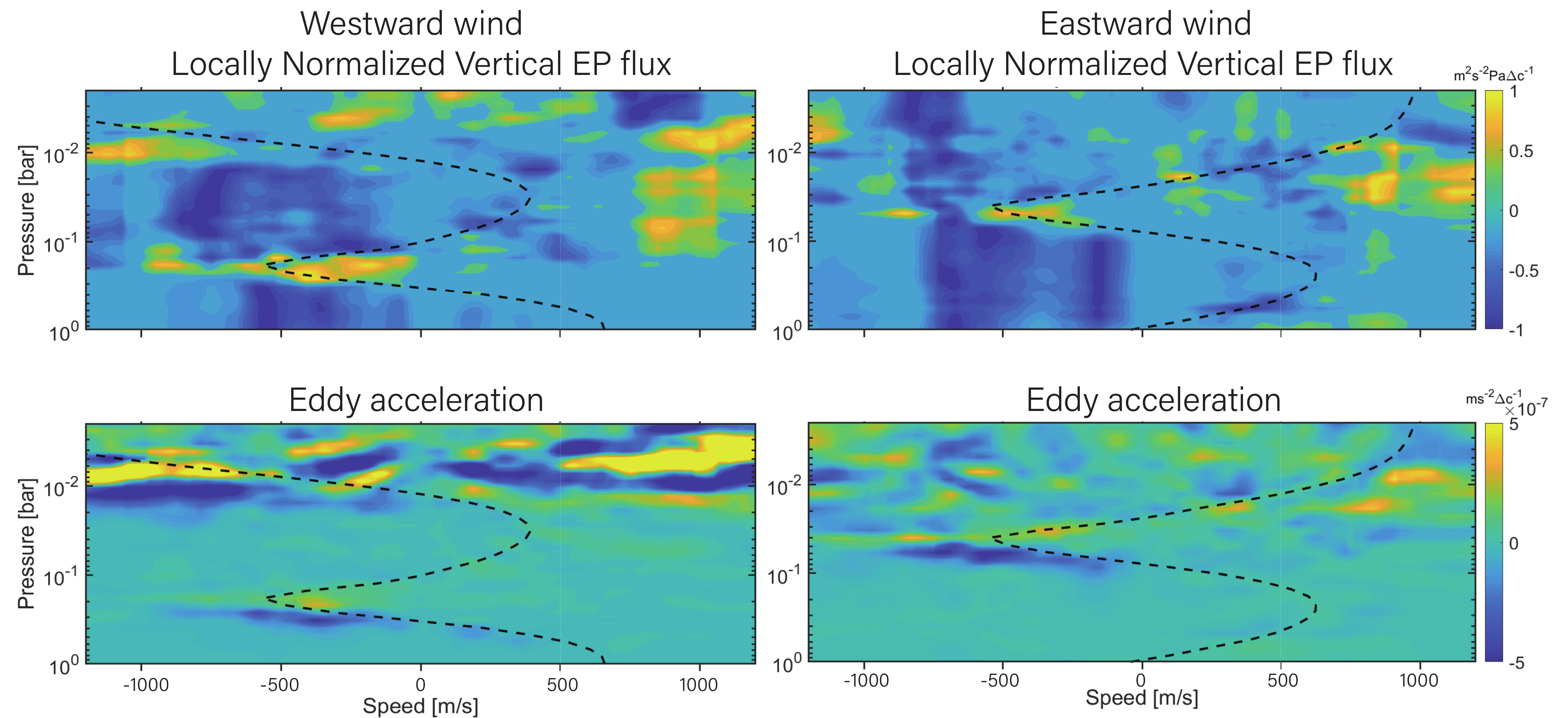}
\caption{ Top: Phasespeed spectra of the Eliassen-Palm flux; Bottom: Eddy acceleration. The thick dashed lines: Equatorial zonal-mean zonal wind for the case of ${\rm In_{strong}Ex_{weak}}$.}
\label{fig:speed_wave}
\end{figure*}

Figure \ref{fig:speed_wave} shows the EP flux and acceleration shades with the zonal-mean zonal wind contours to help us understand the wave absorbing. The spectra show eastward propagating waves with speeds of $\sim 70-120$ ${\rm m}$ ${\rm s^{-1}}$ and westward propagating waves with speeds of$\sim -10- -130$ ${\rm m}$ ${\rm s^{-1}}$, and both waves propagate upward from 1 bar to $10^{-2}$ bar. The westward propagating waves shown here are absorbed at a critical level near the base of the westward jet, 0.3 bar in westward wind cases and 0.08 bar in eastward wind cases, causing westward acceleration on the lower flank (dark blue).
The eastward propagating waves are absorbed at critical levels, 0.06 bar in westward wind cases and 0.02 bar in eastward wind cases, causing eastward acceleration on the lower flank (yellow). The EP fluxes of eastward propagating waves propagate across from 0.2 bar to the top, showing that westward jet is transparent to eastward propagating waves. This figure provides evidence that the quasi-period oscillation is driven by selective absorption of vertically propagating waves.

Note that the quasi-periodic oscillation in our simulations is different from the QBO in the Earth atmosphere. First, the QBO is limited only in the tropics (${\pm 10^\circ}$ latitudes). In contrast, the quasi-periodic oscillation in our simulations expands to $\sim \pm 50$ degrees. It is because tidally locked exoplanets rotate much slower than Earth, and the Rossby deformation radius is much greater (about ${\rm \sim7.4\times10^7}$ ${\rm m}$), causing a wider ``tropics'' and narrower ``extratropics''. For Earth, the source of the equatorial waves originates from tropical convection in the troposphere \citep{holton-1972,bergman-etal-1994}. By contrast, the internal forcing in our simulations is globally isotropic, which may cause tropical waves over a meridionally broader zone. Second, the oscillation period in our model is about twice longer than the QBO period. The QBO period is determined by the magnitude of upward eddy momentum fluxes and the distance between the region of tropical wave sources and the layer of oscillations occurrence \citep{lindzen-etal-1968}. All these are different in our simulations.

We have tested the sensitivity of our simulations to the bottom boundary conditions by lifting the top of perturbation layers from 200 bars to 100 bars. The cases with shallow internal forcing (e.g., ${\rm In_{med}Ex_{strong}Shallow}$) does not show any noticeable changes from that of the control cases. It is likely that the lifting of perturbation layers has little impact on the temporal and horizontal spatial characters of waves. 

\section{Synthetic thermal phase curves}
\label{Lightcurve}

\begin{figure*}
\centering
\includegraphics[width=0.95\textwidth]{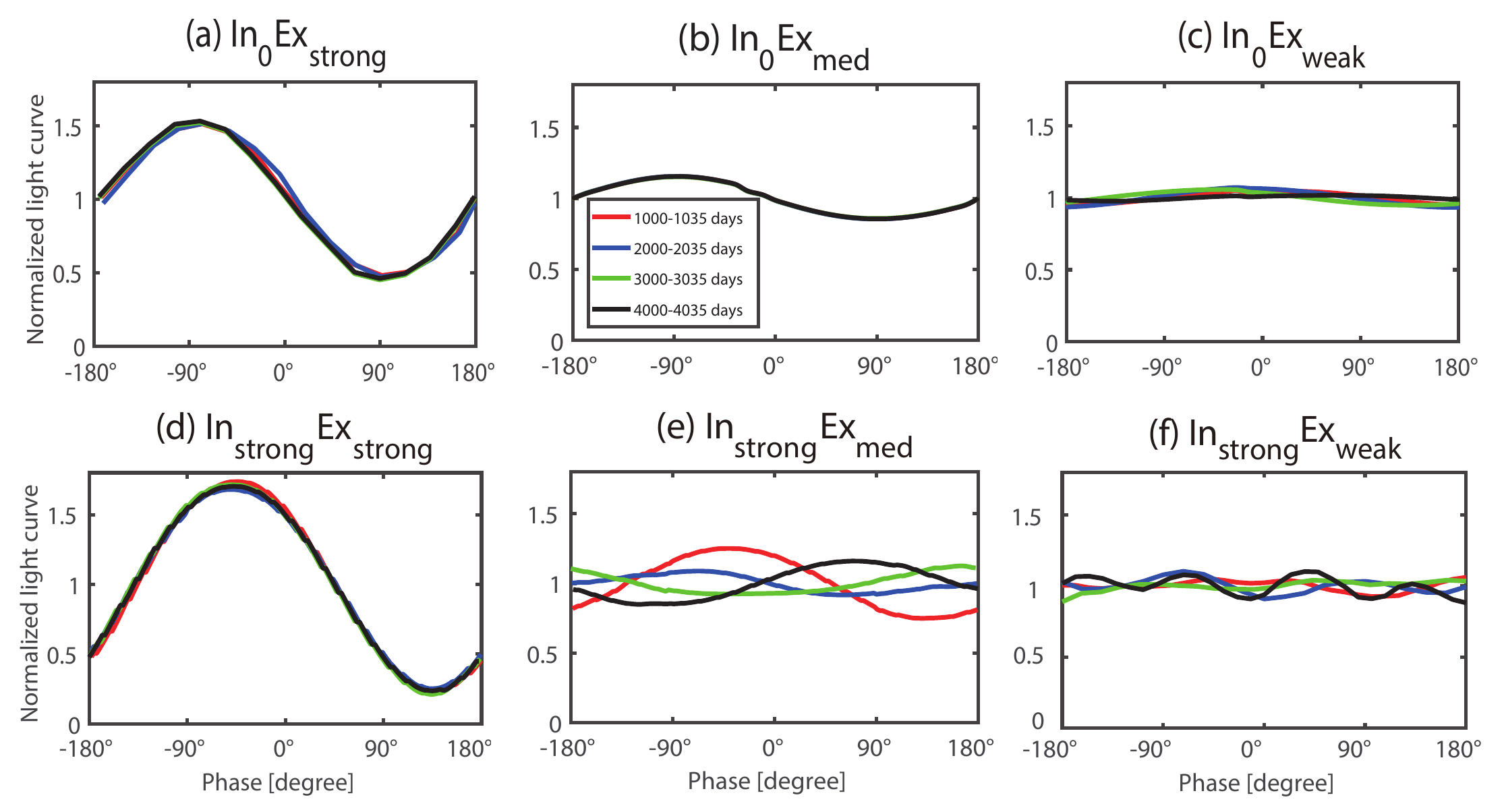}
\caption{ Synthetic thermal phase curves. The thermal fluxes are calculated by integrating radiation fluxes from 50 to 100 mbar, and are normalized by the mean outgoing fluxes. Red curves: integral over 35 days from day 1000 to day 1035, blue curves: integral from day 2000 to day 2035, green curves: integral from day 3000 to day 3035, and black lines: from day 4000 to day 4035. Primary transit occurs at $\pm 180^{\circ}$ and secondary eclipse occurs at $0^{\circ}$.}
\label{fig:12}
\end{figure*}

Heat transports from the dayside to the nightside in the three circulation regimes have significant influences on observational characteristics of giant planets, such as emission spectra and thermal phase curves. Here, we just focus on thermal phase curves. The integral synthetic phase curves are calculated following the method by \citealt{lewis-etal-2010},\citealt{wit-etal-2012}, and \citealt{kataria-etal-2015}. Note that there are no radiative-transfer and cloud schemes in our simulations. The photosphere is around 100 mbar for typical hot Jupiters, according to previous studies \citep{fortney-etal-2006, showman-lewis-etal-2015}. Here, we integrate radiation fluxes from time-average temperatures at layers between 100 mbar and 50 mbar. For each layer, thermal fluxes are calculated using the equations in previous works (Equation (2) in \citet{tan-etal-2020} and Equation (7) in \citet{zhang-etal-2017}). To show the total variations of outgoing radiation fluxes, we plot normalized phase curve variations, which is the quotient of the anomaly of outgoing fluxes and the average of outgoing fluxes.

Figure \ref{fig:12} shows thermal phase curves for six cases with different internal and external forcings. As shown in Figures \ref{fig:12}a, b, and c, amplitudes of the phase curves decrease from 1.7 to 1.1 with decreasing external forcing as the internal forcing is set to zero. This is because of the fact that a stronger external forcing causes larger day-night temperature contrasts. This can also be seen in Figures \ref{fig:12}d, e, and f, in which the internal forcing is strong.

Comparisons between Figures \ref{fig:12}a $\&$ b, b $\&$ e, and c $\&$ f show the effect of internal forcing on thermal phase curves and their dependence on the strength of the internal forcing. When the external forcing is strong (${\rm In_0Ex_{strong}}$ versus ${\rm In_{strong}Ex_{strong}}$), a stronger internal forcing leads to a larger amplitude of the phase curve because the weaker equatorial transports less heats to the nightside (Figure \ref{fig:2}a1 versus Figure \ref{fig:2}a4). For the same reason, the longitudinal shift of the hot spot is smaller. This can be seen from the peaks of the phase curves, showing the hotspot shift, which are at $80^{\circ}$ in the ${\rm In_0Ex_{strong}}$, $70^{\circ}$ in the ${\rm In_0Ex_{med}}$, and $50^{\circ}$  and ${\rm In_{strong}Ex_{strong}}$, respectively. Although internal forcing is strong, the time-dependent internal forcing is not strong enough to cause significant time variability on the phase curve in the case of ${\rm In_{strong}Ex_{strong}}$. When the external forcing is medium (${\rm In_0Ex_{med}}$ versus ${\rm In_{strong}Ex_{med}}$), a strong internal forcing causes time variability of the hot spot shift from $-50^{\circ}$ to $50^{\circ}$. The hot spot shift is associated with the quasi-periodic oscillations, which cause the reversal of heat transports eastward or westward. When the external forcing is weak (${\rm In_0Ex_{weak}}$ versus ${\rm In_{strong}Ex_{weak}}$), a strong internal forcing causes large variations of thermal phase curves, and the sinusoid shape disappears.

The above results indicate that the thermal phase curves of giant planets may be influenced by the internal forcing. It suggests that strong internal forcing could cause variabilities of phase curves, and weak external forcing could cause rather weak phase-curve amplitudes. Previous works have addressed that for brown dwarfs internal forcing can cause large variabilities of thermal phase curves \citep{radigan-etal-2014, yang-etal-2014,zhang-etal-2014,metchev-etal-2015,leggett-etal-2016}. Our work further suggests that small variabilities of thermal phase curves could occur on hot Jupiters if they have strong internal heating.

\section{Discussion and conclusions}
\label{conclusion}
We have showed how internal thermal forcing interacts with external forcing in determining atmospheric circulations of hot Jupiter and strongly irradiated brown dwarfs using a 3D GCM. The major results are:

$\bullet$ When internal forcing is relative weak and external forcing is strong, atmospheric circulation is nearly the same as that in previous works for hot Jupiters, with the equatorial superrotation remained. When both the internal and external forcing are comparably strong, the equatorial eastward jet becomes weaker and moves to lower levels.

$\bullet$ For both weak internal and external forcings, eastward jets develop at middle and high latitudes, while the equatorial eastward jet is weak.

$\bullet$  For strong internal forcing and relatively weak external forcing, quasi-period oscillations of equatorial zonal winds develop at levels above 10 bars, with a quasi-period of about 4 Earth years.

$\bullet$ The changes of circulation patterns largely alter thermal phase curves. When external forcing dominates, large day-night contrast is large, and the hot spot shifts eastward. When both internal and external forcings are strong, the amplitude of thermal phase curves becomes larger and the hot spot shift becomes smaller. The time variability of phase curves increases with increasing internal forcing and decreasing external forcing. For sufficiently strong internal forcing and relatively weak external forcing, the sinusoid shape of phase curves disappears and is not observable. 

The orbital periods of brown dwarfs around white dwarfs may be much shorter than the period we set (Here, the 3.5-earth-day period is a typical value for hot Jupiter). When the rotation period decreases, the jet may be narrower due to the smaller Rossby deformation radius \citep{tan-etal-2020,lee-etal-2020}. It would be interesting to examine the jet strength for faster-rotation brown dwarfs.

In the present study, we assume a frictional drag at the model bottom. The sensitivity of simulations to frictional drag is not tested. \citet{showman-etal-2019} showed that for isolated brown dwarfs and giant planets the banded flow pattern is affected by the strength of the bottom drag. So far, the bottom drag is an unknown parameter. Thus, it is also interesting to examine the sensitivity of atmospheric circulations to bottom drag. 

In our model, we use the Newtonian cooling schemes, instead of using realistic radiative transfer schemes and cloud effects. However, some strong visible absorbers may exist in the atmosphere of hot Jupiters \citep{Hubeny-etal-2003}. More realistic radiative transfer schemes can improve the accuracy of radiative heating and cooling rates and better predict thermal phase curves. We leave this issue for future works. 

High-temperature cloud species, such as TiO and ${\rm MgSiO_3}$ clouds, could form in atmospheres of hot Jupiters and brown dwarfs \citep{spiegel-etal-2009, barstow-etal-2014, helling-casewell-2014}. These clouds may cause cooling throughout scattering and absorption and thus result in different thermal forcings for atmospheric circulations. Clouds may affect simulation results in two aspects: when external forcing dominates, clouds scatter stellar radiation and absorb thermal radiation, which would substantially alter atmospheric circulation patterns and phase curves in both thermal band and visible bands \citep[e.g.][]{lee-etal-2017,parmentier-etal-2016,roman-rauscher-2019}. When internal forcing dominates, on the other hand, cloud radiative effects may cause atmospheric variability and turbulence as well as changes in transient waves, as shown by \citet{tan-showman-2019} and \citet{tan-showman-2021} for isolated brown dwarfs for a typical effective temperature anomaly of hundreds of Kelvins.

\section*{acknowledgements}
The present work was initiated by the late Adam P. Showman. We wish to memorize him with the present paper. We thank Jun Yang for helpful comments. This work is supported by the National Natural Science Foundation of China, under Grants 41888101. Numerical simulation were conducted at the High-performance Computing Platform of Peking University.

\bibliography{yuchen_paper}

\begin{thebibliography}{}
\expandafter\ifx\csname natexlab\endcsname\relax\def\natexlab#1{#1}\fi
\providecommand{\url}[1]{\href{#1}{#1}}
\providecommand{\dodoi}[1]{doi:~\href{http://doi.org/#1}{\nolinkurl{#1}}}
\providecommand{\doeprint}[1]{\href{http://ascl.net/#1}{\nolinkurl{http://ascl.net/#1}}}
\providecommand{\doarXiv}[1]{\href{https://arxiv.org/abs/#1}{\nolinkurl{https://arxiv.org/abs/#1}}}

\bibitem[{Andrews {et~al.}(1987)Andrews, Holton, \& Leovy}]{andrews-1987}
Andrews, D., Holton, J., \& Leovy, C. 1987, Middle Atmosphere Dynamics,
  Vol.~40, 489

\bibitem[{Artigau {et~al.}(2009)Artigau, Bouchard, Doyon, \&
  Lafreniere}]{artigau-etal-2009}
Artigau, E., Bouchard, S., Doyon, R., \& Lafreniere, D. 2009, The Astrophysical
  Journal, 701, 1534

\bibitem[{Baldwin {et~al.}(2001)Baldwin, Gray, Dunkerton, Hamilton, Haynes,
  Randel, Holton, Alexander, Hirota, Horinouchi, {et~al.}}]{baldwin-etal-2001}
Baldwin, M.~P., Gray, L.~J., Dunkerton, T.~J., {et~al.} 2001, Reviews of
  Geophysics, 39, 179

\bibitem[{Baraffe {et~al.}(2009)Baraffe, Chabrier, \&
  Barman}]{baraffe-etal-2009}
Baraffe, I., Chabrier, G., \& Barman, T. 2009, Reports on Progress in Physics,
  73

\bibitem[{Barstow {et~al.}(2014)Barstow, Aigrain, Irwin, Hackler, \&
  Fletcher}]{barstow-etal-2014}
Barstow, J.~K., Aigrain, S., Irwin, P. G.~J., Hackler, T., \& Fletcher, L.~N.
  2014, The Astrophysical Journal, 786, 154

\bibitem[{Bayliss {et~al.}(2016)Bayliss, Hojjatpanah, Santerne, Dragomir, Zhou,
  Shporer, Colon, Almenara, Armstrong, Barrado, Barros, Bento, Boisse, Bouchy,
  Brown, Cameron, Cochran, Demangeon, \& Tsantaki}]{bayliss-etal-2016}
Bayliss, D., Hojjatpanah, S., Santerne, A., {et~al.} 2016, The Astronomical
  Journal, 153, 15

\bibitem[{Bergman \& Salby(1994)}]{bergman-etal-1994}
Bergman, J.~W., \& Salby, M.~L. 1994, Journal of the Atmospheric Sciences, 51,
  3791

\bibitem[{Burrows {et~al.}(2001)Burrows, Hubbard, Lunine, \&
  Liebert}]{burrows-etal-2001}
Burrows, A., Hubbard, W.~B., Lunine, J.~I., \& Liebert, J. 2001, Reviews of
  Modern Physics, 73, 719

\bibitem[{Casewell {et~al.}(2012)Casewell, Burleigh, Wynn, Alexander,
  Napiwotzki, Lawrie, Dobbie, Jameson, \& Hodgkin}]{casewell-etal-2012}
Casewell, S.~L., Burleigh, M.~R., Wynn, G.~A., {et~al.} 2012, The Astrophysical
  Journal, 759, 1

\bibitem[{Chao \& Schoeberl(1984)}]{chao-etal-1984}
Chao, W.~C., \& Schoeberl, M.~R. 1984, Journal of the Atmospheric Sciences, 41,
  1893

\bibitem[{Charney(1947)}]{charney-1947}
Charney, J.~G. 1947, Journal of Meteorology, 4, 136

\bibitem[{Cooper \& Showman(2005)}]{cooper-etal-2005}
Cooper, C.~S., \& Showman, A.~P. 2005, Astrophysical Journal, 629, L45

\bibitem[{Cushing {et~al.}(2016)Cushing, Hardegreeullman, Trucks, Morley,
  Gizis, Marley, Fortney, Kirkpatrick, Gelino, Mace,
  {et~al.}}]{cushing-etal-2016}
Cushing, M.~C., Hardegreeullman, K.~K., Trucks, J.~L., {et~al.} 2016, The
  Astrophysical Journal, 823, 152

\bibitem[{De~Wit {et~al.}(2012)De~Wit, Gillon, Demory, \&
  Seager}]{wit-etal-2012}
De~Wit, J., Gillon, M., Demory, B., \& Seager, S. 2012, Astronomy and
  Astrophysics, 548, 128

\bibitem[{Debras {et~al.}(2020)Debras, Mayne, Baraffe, Jaupart, Mourier, Laibe,
  Goffrey, \& Thuburn}]{debras-etal-2020}
Debras, F., Mayne, N.~J., Baraffe, I., {et~al.} 2020, Astronomy and
  Astrophysics, 633

\bibitem[{Espinoza \& Jordan(2015)}]{espinoza-etal-2015}
Espinoza, N., \& Jordan, A. 2015, Monthly Notices of the Royal Astronomical
  Society, 450, 1879

\bibitem[{Farrell \& Ioannou(2008)}]{farrell-etal-2008}
Farrell, B.~F., \& Ioannou, P.~J. 2008, Journal of the Atmosphere Sciences, 65,
  3353

\bibitem[{Fortney {et~al.}(2006)Fortney, Cooper, Showman, Marley, \&
  Freedman}]{fortney-etal-2006}
Fortney, J.~J., Cooper, C.~S., Showman, A.~P., Marley, M.~S., \& Freedman,
  R.~S. 2006, the Astrophysical Journal, 652, 746

\bibitem[{Fortney \& Nettelmann(2010)}]{fortney-etal-2010}
Fortney, J.~J., \& Nettelmann, N. 2010, Space Science Reviews, 152, 423

\bibitem[{Freytag {et~al.}(2010)Freytag, Allard, Ludwig, Homeier, \&
  Steffen}]{freytag-etal-2010}
Freytag, B., Allard, F., Ludwig, H.-G., Homeier, D., \& Steffen, M. 2010,
  Astronomy and Astrophisics, 513, 14

\bibitem[{Fritts(1984)}]{fritts-1984}
Fritts, D.~C. 1984, Reviews of Geophysics and Space Physics, 22

\bibitem[{Guillot \& Showman(2002)}]{guillot-etal-2002}
Guillot, T., \& Showman, A.~P. 2002, Astronomy and Astrophysics, 385, 156

\bibitem[{Hammond \& Pierrehumbert(2018)}]{hammond-etal-2018}
Hammond, M., \& Pierrehumbert, R.~T. 2018, The Astrophysical Journal, 869, 65

\bibitem[{Helling \& Casewell(2014)}]{helling-casewell-2014}
Helling, C., \& Casewell, S. 2014, The Astronomy and Astrophysics Review, 22,
  80

\bibitem[{Heng \& Showman(2015)}]{heng-showman-2015}
Heng, K., \& Showman, A.~P. 2015, Annual Review of Earth and Planetary
  Sciences, 43, 509

\bibitem[{Holton(1972)}]{holton-1972}
Holton, J.~R. 1972, Journal of the Atmospheric Sciences, 29, 368

\bibitem[{Holton(2004)}]{holton-2004}
---. 2004, An Introduction to Dynamic Meteorology, Fourth Edition (Academic
  Press)

\bibitem[{Hu \& Ding(2013)}]{hu-etal-2013}
Hu, Y., \& Ding, F. 2013, Scientia Sinica, 43, 1356

\bibitem[{Hubeny {et~al.}(2003)Hubeny, Burrows, \& Sudarsky}]{Hubeny-etal-2003}
Hubeny, I., Burrows, A., \& Sudarsky, D. 2003, The Astrophysical Journal, 594,
  1011

\bibitem[{Iro {et~al.}(2005)Iro, Bezard, \& Guillot}]{iro-etal-2005}
Iro, N., Bezard, B., \& Guillot, T. 2005, Astronomy and Astrophysics, 436, 719

\bibitem[{Jackson {et~al.}(2019)Jackson, Adams, Sandidge, Kreyche, \&
  Briggs}]{jackson-etal-2019}
Jackson, B., Adams, E., Sandidge, W., Kreyche, S., \& Briggs, J. 2019, The
  Astronomical Journal, 157, 239

\bibitem[{Kataria {et~al.}(2015)Kataria, Showman, Fortney, Stevenson, Line,
  Kreidberg, Bean, \& Désert}]{kataria-etal-2015}
Kataria, T., Showman, A.~P., Fortney, J.~J., {et~al.} 2015, Astrophysical
  Journal, 801, 86

\bibitem[{Knutson {et~al.}(2008)Knutson, Charbonneau, Allen, Burrows, \&
  Megeath}]{knutson-etal-2008}
Knutson, H.~A., Charbonneau, D., Allen, L.~E., Burrows, A., \& Megeath, S.~T.
  2008, The Astrophysical Journal, 673, 526

\bibitem[{Knutson {et~al.}(2007)Knutson, David, Allen, Fortney, Eric, Cowan,
  Showman, Cooper, \& S~Thomas}]{knutson-etal-2007}
Knutson, H.~A., David, C., Allen, L.~E., {et~al.} 2007, Nature, 447, 183

\bibitem[{Komacek \& Showman(2016)}]{komacek-etal-2016}
Komacek, T.~D., \& Showman, A.~P. 2016, The Astrophysical Journal, 821, 16

\bibitem[{Lee {et~al.}(2020)Lee, Casewell, Chubb, Hammond, Tan, Tsai, \&
  Pierrehumbert}]{lee-etal-2020}
Lee, G. K.~H., Casewell, S.~L., Chubb, K.~L., {et~al.} 2020, Monthly Notices of
  the Royal Astronomical Society, 000, 1

\bibitem[{Lee {et~al.}(2017)Lee, Wood, Dobbs-Dixon, Rice, \&
  Helling}]{lee-etal-2017}
Lee, G. K.~H., Wood, K., Dobbs-Dixon, I., Rice, A., \& Helling, C. 2017,
  Astronomy and Astrophysics, 601, A22

\bibitem[{Leggett {et~al.}(2016)Leggett, Cushing, Hardegree-Ullman, Trucks,
  Marley, Morley, Saumon, Carey, Fortney, Gelino, Gizis, Kirkpatrick, \&
  Mace}]{leggett-etal-2016}
Leggett, S., Cushing, M., Hardegree-Ullman, K., {et~al.} 2016, The
  Astrophysical Journal, 830

\bibitem[{Lewis {et~al.}(2010)Lewis, Showman, Fortney, Marley, Freedman, \&
  Lodders}]{lewis-etal-2010}
Lewis, N.~K., Showman, A.~P., Fortney, J.~J., {et~al.} 2010, Astrophysical
  Journal, 720, 344

\bibitem[{Lindzen(1970)}]{lindzen-etal-1970}
Lindzen, R.~S. 1970, Journal of the Atmospheric Sciences, 27, 394

\bibitem[{Lindzen(1971)}]{lindzen-etal-1971}
---. 1971, Journal of the Atmospheric Sciences, 28, 1452

\bibitem[{Lindzen \& Holton(1968)}]{lindzen-etal-1968}
Lindzen, R.~S., \& Holton, J.~R. 1968, Journal of the Atmospheric Sciences, 25,
  1095

\bibitem[{Liu \& Showman(2012)}]{liu-etal-2012}
Liu, B., \& Showman, A.~P. 2012, Astrophysical Journal, 770, 311

\bibitem[{Marley \& Robinson(2015)}]{marley-etal-2015}
Marley, M.~S., \& Robinson, T.~D. 2015, Annual Review Of Astronomy And
  Astrophysics, 53, 279

\bibitem[{Mayne {et~al.}(2017)Mayne, Debras, Baraffe, Thuburn, Amundsen,
  Acreman, Smith, Browning, Manners, \& Wood}]{mayne-etal-2017}
Mayne, N.~J., Debras, F., Baraffe, I., {et~al.} 2017, Astronomy and
  Astrophysics, 604, 27

\bibitem[{Mendonça(2020)}]{mendonca-2020}
Mendonça, J.~M. 2020, Monthly Notices of the Royal Astronomical Society, 491,
  1456

\bibitem[{Metchev {et~al.}(2015)Metchev, Heinze, Apai, Flateau, Radigan,
  Burgasser, Marley, Artigau, Plavchan, \& Goldman}]{metchev-etal-2015}
Metchev, S., Heinze, A., Apai, D., {et~al.} 2015, The Astrophysical Journal,
  799, 154

\bibitem[{Parmentier {et~al.}(2016)Parmentier, Fortney, Showman, Morley, \&
  Marley}]{parmentier-etal-2016}
Parmentier, V., Fortney, J.~J., Showman, A.~P., Morley, C., \& Marley, M.~S.
  2016, The Astrophysical Journal, 828, 22

\bibitem[{Perez-Becker \& Showman(2013)}]{perez-etal-2013}
Perez-Becker, D., \& Showman, A.~P. 2013, The Astrophysical Journal, 776, 134

\bibitem[{Radigan {et~al.}(2014)Radigan, Lafreniere, Jayawardhana, \&
  Artigau}]{radigan-etal-2014}
Radigan, J., Lafreniere, D., Jayawardhana, R., \& Artigau, E. 2014, The
  Astrophysical Journal, 793, 75

\bibitem[{Roman \& Rauscher(2019)}]{roman-rauscher-2019}
Roman, M., \& Rauscher, E. 2019, The Astrophysical Journal, 872, 16

\bibitem[{Scott \& Polvani(2007)}]{scott-etal-2007}
Scott, R.~K., \& Polvani, L.~M. 2007, Journal of the Atmospheric Sciences, 64,
  3158

\bibitem[{Showman {et~al.}(2008{\natexlab{a}})Showman, Cooper, Fortney, \&
  Marley}]{showman-etal-2008}
Showman, A.~P., Cooper, C.~S., Fortney, J.~J., \& Marley, M.~S.
  2008{\natexlab{a}}, Astrophysical Journal, 682, 559

\bibitem[{Showman {et~al.}(2008{\natexlab{b}})Showman, Fortney, Lian, Marley,
  Freedman, Knutson, \& Charbonneau}]{showman-fortney-etal-2008}
Showman, A.~P., Fortney, J.~J., Lian, Y., {et~al.} 2008{\natexlab{b}},
  Astrophysical Journal, 699, 564

\bibitem[{Showman \& Guillot(2002)}]{showman-etal-2002}
Showman, A.~P., \& Guillot, T. 2002, Astronomy and Astrophysics, 385, 166

\bibitem[{Showman \& Kaspi(2013)}]{showman-etal-2013}
Showman, A.~P., \& Kaspi, Y. 2013, Astrophysical Journal, 776, 85

\bibitem[{Showman {et~al.}(2015)Showman, Lewis, \&
  Fortney}]{showman-lewis-etal-2015}
Showman, A.~P., Lewis, N.~K., \& Fortney, J.~J. 2015, Physics, 801, 1816

\bibitem[{Showman \& Polvani(2011)}]{showman-etal-2011}
Showman, A.~P., \& Polvani, L.~M. 2011, Astrophysical Journal, 738, 71

\bibitem[{Showman {et~al.}(2020)Showman, Tan, \&
  Parmentier}]{showman2020review}
Showman, A.~P., Tan, X., \& Parmentier, V. 2020, Space Science Reviews, 216,
  139

\bibitem[{Showman {et~al.}(2019)Showman, Tan, \& Zhang}]{showman-etal-2019}
Showman, A.~P., Tan, X., \& Zhang, X. 2019, The Astrophysical Journal, 883, 4

\bibitem[{Siverd {et~al.}(2012)Siverd, Beatty, Pepper, Eastman, Collins,
  Bieryla, Latham, Buchhave, Jensen, Crepp, {et~al.}}]{siverd-etal-2012}
Siverd, R.~J., Beatty, T.~G., Pepper, J., {et~al.} 2012, The Astrophysical
  Journal, 761, 123

\bibitem[{Spiegel {et~al.}(2009)Spiegel, Silverio, \&
  Burrows}]{spiegel-etal-2009}
Spiegel, D.~S., Silverio, K., \& Burrows, A. 2009, The Astrophysical Journal,
  699, 1487

\bibitem[{Tan \& Showman(2019)}]{tan-showman-2019}
Tan, X., \& Showman, A.~P. 2019, The Astrophysical Journal, 874, 111

\bibitem[{{Tan} \& {Showman}(2020)}]{tan-etal-2020}
{Tan}, X., \& {Showman}, A.~P. 2020, arXiv e-prints, arXiv:2001.06269.
\newblock \doarXiv{2001.06269}

\bibitem[{Tan \& Showman(2021)}]{tan-showman-2021}
Tan, X., \& Showman, A.~P. 2021, Monthly Notices of the Royal Astronomical
  Society, 502, 678

\bibitem[{Thorngren {et~al.}(2019)Thorngren, Gao, \&
  Fortney}]{thorngren-etal-2019}
Thorngren, D., Gao, P., \& Fortney, J.~J. 2019, The Astrophysical Journal, 884

\bibitem[{Thrastarson \& Cho(2010)}]{thrastarson-etal-2010}
Thrastarson, H.~T., \& Cho, J.~Y. 2010, The Astrophysical Journal, 716, 144

\bibitem[{Tsai {et~al.}(2014)Tsai, Ian, \& Gu}]{tsai-etal-2014}
Tsai, S.~M., Ian, D.-D., \& Gu, P.~G. 2014, Astrophysical Journal, 793

\bibitem[{Vanzandt \& Fritts(1989)}]{vanzandt-etal-1989}
Vanzandt, T.~E., \& Fritts, D.~C. 1989, Pure and Applied Geophysics, 130, 399

\bibitem[{Wang \& Yang(2020)}]{wang-yang-2020}
Wang, S., \& Yang, J. 2020, The Astrophysical Journal, 907, 13

\bibitem[{Wilson {et~al.}(2014)Wilson, Rajan, \& Patience}]{wilson-etal-2014}
Wilson, P.~A., Rajan, A., \& Patience, J. 2014, Astronomy and Astrophysics,
  566, 111

\bibitem[{Yang {et~al.}(2014)Yang, Apai, Marley, Saumon, Morley, Buenzli,
  Artigau, Radigan, Metchev, Burgasser, Mohanty, Lowrance, Showman, Karalidi,
  Flateau, \& Heinze}]{yang-etal-2014}
Yang, H., Apai, D., Marley, M.~S., {et~al.} 2014, The Astrophysical Journal,
  798,

\bibitem[{Yang \& Yu(2016)}]{yang-yu-2016}
Yang, M., \& Yu, Y. 2016, Climate Dynamics, 47, 1943–1959

\bibitem[{Youdin \& Mitchell(2010)}]{youdin-mitchell-2010}
Youdin, A.~N., \& Mitchell, J.~L. 2010, The Astrophysical Journal, 721, 1113

\bibitem[{Zhang \& Showman(2014)}]{zhang-etal-2014}
Zhang, X., \& Showman, A.~P. 2014, Astrophysical Journal, 788,

\bibitem[{Zhang \& Showman(2017)}]{zhang-etal-2017}
---. 2017, The Astrophysical Journal, 836, 73

\end{thebibliography}
\bibliographystyle{aasjournal}
\end{document}